\documentclass[12pt]{article}
\usepackage{latexsym}
\usepackage{graphicx}% Include figure files
\title{Quantum mechanics: Myths and facts}
\author{Hrvoje Nikoli\'c \\
Theoretical Physics Division, Rudjer Bo\v{s}kovi\'{c} Institute, \\
P.O.B. 180, HR-10002 Zagreb, Croatia \\
{\normalsize e-mail: hrvoje@thphys.irb.hr} \\
\makebox[1in]{} \\
}
\date{\today}
\begin{document}
\maketitle
\begin{abstract}
A common understanding of quantum mechanics (QM) among students and 
practical users is often plagued by a number of ``myths", that is,  
widely accepted claims on which there is not really a general consensus
among experts in foundations of QM. These myths include 
wave-particle duality, time-energy uncertainty relation, 
fundamental randomness, the absence of measurement-independent reality, 
locality of QM, nonlocality of QM, 
the existence of well-defined relativistic QM, the claims that 
quantum field theory (QFT) solves the problems of relativistic QM or
that QFT is a theory of particles, as well as myths on black-hole entropy.  
The fact is that the existence of various theoretical and interpretational 
ambiguities underlying these myths does not yet allow us to accept
them as proven facts. I review the main arguments and counterarguments 
lying behind these myths and conclude that QM is still a 
not-yet-completely-understood theory open to further fundamental 
research. 
\end{abstract}

\tableofcontents

\section{Introduction}

On the technical level, quantum mechanics (QM) is a set of mathematically 
formulated prescriptions that serve for calculations of probabilities 
of different measurement outcomes. The calculated probabilities
agree with experiments. This is the fact! 
From a pragmatic point of view, this is also enough.
Pragmatic physicists are interested only in 
these pragmatic aspects of QM, which is fine. Nevertheless, many 
physicists are not only interested in the pragmatic aspects, 
but also want to understand nature on a deeper conceptual level. Besides, 
a deeper understanding of nature on the conceptual level may also induce 
a new development of pragmatic aspects. Thus, the conceptual
understanding of physical phenomena is also an important aspect of physics.
Unfortunately, the conceptual issues turn out to be particularly difficult in 
the most fundamental physical theory currently known -- quantum theory.
     
Textbooks on QM usually emphasize the pragmatic technical aspects, 
while the discussions of the conceptual issues are usually 
avoided or reduced to simple authoritative claims without a detailed 
discussion. This causes a common (but wrong!) impression among physicists 
that all conceptual problems of QM are already solved 
or that the unsolved problems are not really physical (but rather
``philosophical"). The purpose of the present paper is to 
warn students, teachers, and practitioners that some of the 
authoritative claims on conceptual aspects of QM that they often heard 
or read may be actually wrong, that a certain number of serious 
physicists still copes with these foundational aspects of QM, 
and that there is not yet a general consensus among experts on 
answers to some of the most fundamental questions.
To emphasize that some widely accepted authoritative claims on QM
are not really proven, I refer to them as ``myths".
In the paper, I review the main facts that support
these myths, but also explain why these facts do not really prove the myths
and review the main alternatives.
The paper is organized such that each section is devoted 
to another myth, while the title of each section carries 
the basic claim of the corresponding myth. (An exception 
is the concluding section where I attempt to identify 
the common origin of all these myths.)
The sections are 
roughly organized from more elementary myths  
towards more advanced ones,
but they do not necessarily need to be read in that order. 
The style of presentation is adjusted to readers who are 
already familiar with the technical aspects of QM, but 
want to complete their knowledge with a better understanding 
of the conceptual issues. Nevertheless,
the paper is attempted to be very pedagogical and readable 
by a wide nonexpert audience. However,
to keep the balance and readibility by a wide physics audience,
with a risk of making the paper less pedagogical,
in some places I am forced to omit some technical details
(especially in the most advanced sections, Secs.~\ref{QFTP} and \ref{BH}),
keeping only those conceptual and technical details that are essential
for understanding why some myths are believed to be true and why
they may not be so. Readers interested in more technical details
will find them in more specialized cited references, many of which
are pedagogically oriented reviews.

As a disclaimer, it is also fair to stress that
although this paper is intended to be a review of 
different views and interpretations of various aspects of QM,
it is certainly not completely neutral and unbiased. Some 
points of view are certainly emphasized more than the others,
while some are not even mentioned, 
reflecting subjective preferences of the author.
Moreover, the reader does not necessarily need to agree with
all conclusions and suggestions presented in this paper, as they also 
express a subjective opinion of the author,
which, of course, is also open to further criticism.
By dissolving various myths in QM, it is certainly not intention
of the author to create new ones. 
The main intention of the author
is to provoke new thinking of the reader
about various aspects of QM that previously might have been taken 
by him/her for granted, not necessarily to convince the reader
that views presented here are the right ones.   
Even the claims that are proclaimed as ``facts" in this paper
may be questioned by a critical reader.
It also cannot be overemphasized that ``myths" in this paper
do not necessarily refer to claims that are wrong, but merely
to claims about which there is not yet a true consensus.

\section{In QM, there is a wave-particle duality}

\subsection{Wave-particle duality as a myth}

In introductory textbooks on QM, as well as in popular texts on 
QM, a conceptually strange character of QM is often verbalized 
in terms of {\it wave-particle duality}. According to this duality, 
fundamental microscopic objects such as electrons and photons 
are neither pure particles nor pure waves, but both waves and particles.
Or more precisely, in some conditions they behave as waves, 
while in other conditions they behave as particles.
However, in more advanced and technical textbooks on QM, 
the wave-particle duality is rarely mentioned. Instead, 
such serious textbooks talk only about waves, i.e., wave functions 
$\psi({\bf x},t)$. The waves do not need to be plane waves 
of the form $\psi({\bf x},t)=e^{i({\bf kx}-\omega t)}$, but, 
in general, may have an arbitrary dependence on ${\bf x}$ and 
$t$. At time $t$, the wave can be said to behave as a particle
if, at that time, the wave is {\em localized} 
around a single value of ${\bf x}$.
In the ideal case, if 
\begin{equation}\label{x}
\psi({\bf x})=\sqrt{\delta^3({\bf x}-{\bf x}')} ,
\end{equation}
then the position ${\bf x}$ of the particle has a definite 
value ${\bf x}'$.
The state (\ref{x}) is the eigenstate of the position operator, 
with the eigenvalue ${\bf x}'$. 
Typically, the wave attains such a localized-particle shape through
a wave-function collapse associated with a measurement of a 
particle position. Moreover, the wave may appear as a pointlike
particle for a long time if the particle position is measured
many times in sequence with a small time interval between two 
measurements. This makes the wave to appear as a classical particle 
with a trajectory, which occurs, e.g., in cloud chambers.  
However, the position operator 
is just one of many (actually, infinitely many) hermitian operators
in QM. Each hermitian operator corresponds to an observable, and
it is widely accepted (which, as we shall see later, is also one of 
the myths) that the position operator does not enjoy any privileged 
role. From that, widely accepted, point of view, there is nothing 
dual about QM; electrons and photons {\em always} behave as 
waves, while a particlelike behavior corresponds only to a 
special case (\ref{x}). In this sense, the wave-particle duality is 
nothing but a myth.

But why then the wave-particle duality is so often mentioned?
One reason is philosophical; the word ``duality" sounds 
very ``deep" and ``mysterious" from a philosophical point of view, 
and some physicists obviously like it, despite the fact 
that a dual picture is not supported by the usual technical 
formulation of QM.  
Another reason is historical; in early days of QM, 
it was an experimental fact that electrons and photons 
sometimes behave as particles and sometimes as waves,
so a dual interpretation was perhaps natural at that time
when quantum theory was not yet well understood.    

From above, one may conclude that the notion of ``wave-particle duality"
should be completely removed from a modern talk on QM. 
However, this is not necessarily so. Such a concept may still 
make sense if interpreted in a significantly different way. 
One way is purely linguistic; it is actually common to say that 
electrons and photons are ``particles", having in mind that 
the word ``particle" has a very different meaning than the same 
word in classical physics. In this sense, electrons and photons 
are both ``particles" (because we call them so) and ``waves"
(because that is what, according to the usual interpretation, 
they really are). 

Another meaningful way of retaining the notion of 
``wave-particle duality" is to understand it as a
quantum-classical duality, becuse each classical theory
has the corresponding quantum theory, and vice versa. 
However, the word ``duality" is not the best word for this 
correspondence, because the corresponding quantum and classical theories 
do not enjoy the same rights. Instead, the classical theories 
are merely approximations of the quantum ones.  
 
\subsection{Can wave-particle duality be taken seriously?}

However, is it possible that the ``wave-particle duality" has a 
literal meaning; that, in some sense, electrons and photons 
really {\em are} both particles and waves? Most experts 
for foundations of QM will probably say -- no! Nevertheless, 
such a definite ``no" is also an unproved myth.
Of course, such a definite ``no" is correct if it refers only 
to the usual formulation of QM. But who says that the usual 
formulation of QM is the ultimate theory that will never be 
superseded by an even better theory? (A good scientist will 
never say that for any theory.) 
In fact, such a modification of the usual quantum theory already 
exists. I will refer to it as the {\em Bohmian} interpretation 
of QM \cite{bohm}, but it is also known under the names 
``de Broglie-Bohm" interpretation and ``pilot-wave" interpretation.
(For recent pedagogic expositions of this interpretation, see
\cite{tumul,pas}, for a pedagogic comparison with other formulations 
of QM, see \cite{mnogoAJP}, and for un unbiased review 
of advantages and disadvantages of this interpretation,
see \cite{pas2}.)
This interpretation consists of {\em two} equations. One is the 
standard Schr\"odinger equation that describes the wave-aspect
of the theory, while the other is a classical-like equation that 
describes a particle trajectory. The equation for the 
trajectory is such that the force on the particle depends on the 
wave function, so that the motion of the particle differs 
from that in classical physics, which, in turn, can be used to explain
all (otherwise strange) quantum phenomena. In this interpretation, 
{\em both} 
the wave function and the particle position are fundamental 
entities. If any known interpretation of QM respects 
a kind of wave-particle duality, then it is the Bohmian interpretation.
More on this interpretation (which also provides a counterexample 
to some other myths of QM) will be presented in subsequent sections.   

\section{In QM, there is a time-energy uncertainty relation}

\subsection{The origin of a time-energy uncertainty relation}

For simplicity, consider a particle moving in one dimension.
In QM, operators corresponding to the position $x$ and the momentum $p$ satisfy 
the commutation relation
\begin{equation}\label{comrelxp}
[\hat{x},\hat{p}]=i\hbar ,
\end{equation}
where $[A,B]\equiv AB-BA$.
As is well known, this commutation relation implies the 
position-momentum Heisenberg uncertainty relation
\begin{equation}\label{Hxp}
\Delta x \Delta p \geq \frac{\hbar}{2} .
\end{equation}
It means that one cannot measure both the particle momentum and 
the particle position with arbitrary accuracy. For example, 
the wave function correponding to a definite momentum 
is an eigenstate of the momentum operator
\begin{equation}\label{momop}
\hat{p}=-i\hbar\frac{\partial}{\partial x} .
\end{equation}
It is easy to see that such a wave function must be proportional
to a plane wave $e^{ipx/\hbar}$. 
On the other hand, the wave function corresponding 
to an eigenstate of the position operator is essentially a $\delta$-function 
(see (\ref{x})). It is clear that a wave function cannot be 
both a plane wave and a $\delta$-function, which, in the usual
formulation of QM, explains why one cannot measure both  
the momentum and the position with perfect accuracy.

There is a certain analogy between the couple position-momentum 
and the couple time-energy. In particular, a wave function
that describes a particle with a definite energy $E$ is proportional to
a plane wave $e^{-iEt/\hbar}$. Analogously, one may imagine that 
a wave function corresponding to a definite time is essentially 
a $\delta$-function in time. In analogy with (\ref{Hxp}),
this represents an essence of the reason 
for writing the time-energy uncertainty relation
\begin{equation}\label{HtE}
\Delta t \Delta E \geq \frac{\hbar}{2} .
\end{equation}   
In introductory textbooks on QM, 
as well as in popular texts on QM, the time-energy uncertainty relation
(\ref{HtE}) is often presented as a fact enjoying the same rights 
as the position-momentum uncertainty relation (\ref{Hxp}). 
Nevertheless, there is a great difference between these two uncertainty
relations. Whereas the position-momentum uncertainty relation (\ref{Hxp}) 
is a fact, the time-energy uncertainty relation (\ref{HtE}) is a myth!

\subsection{The time-energy uncertainty relation is not fundamental}

Where does this difference come from? The main difference lies in the fact 
that energy is {\em not} represented by an 
operator analogous to (\ref{momop}), i.e., energy is not represented 
by the operator $i\hbar\partial/\partial t$. Instead, energy is represented
by a much more complicated operator called Hamiltonian, usually having 
the form
\begin{equation}
\hat{H}=\frac{\hat{p}^2}{2m}+V(\hat{x}) .
\end{equation} 
Nothing forbids the state $\psi(x,t)$ 
to be an eigenstate of $\hat{H}$ at a definite value of $t$.
This difference has a deeper origin in the fundamental postulates 
of QM, according to which quantum operators 
are operators on the space of functions depending on $x$, {\em not} on the 
space of functions depending on $t$. Thus, space and time have very different 
roles in nonrelativistic QM. While $x$ is an operator, $t$ is only 
a classical-like parameter. 
A total probability that must be equal to 1 
is an integral of the form
\begin{equation}\label{integ1}
\int_{-\infty}^{\infty} dx\, \psi^*(x,t)\psi(x,t) ,
\end{equation}
not an integral of the form
\begin{equation}\label{integ2}
\int_{-\infty}^{\infty} \int_{-\infty}^{\infty} dx\, dt\, \psi^*(x,t)\psi(x,t) .
\end{equation}
In fact, if $\psi(x,t)$ is a solution of the Schr\"odinger equation,
then, when the integral (\ref{integ1}) 
is finite, the integral (\ref{integ2}) 
is not finite. An analogous statement is also true for one or more particles 
moving in 3 dimensions; the probability density $\psi^*\psi$ is not 
to be integrated over time.

As the time $t$ is not an operator in QM, a commutation 
relation analogous to (\ref{comrelxp}) but with the replacements
$x\rightarrow t$, $p\rightarrow H$, does not make sense. This is another 
reason why the time-energy uncertainty relation (\ref{HtE}) is not 
really valid in QM. Nevertheless, there are attempts to replace
the parameter $t$ with another quantity $T$, so that an analog of  
(\ref{comrelxp})
\begin{equation}\label{comrelTH}
[\hat{T},\hat{H}]=-i\hbar 
\end{equation}
is valid. However, there is a theorem due to Pauli that says that this is 
impossible \cite{pauli}. The two main assumptions of the Pauli theorem 
are that $T$ and $H$ must be hermitian operators (because only hermitian
operators have real eigenvalues corresponding to real physical 
quantities) and that the spectrum of $H$ must be bounded from 
below (which corresponds to the physical requirement 
that energy should not have the possibility to become arbitrarily negative, 
because otherwise such a system would not be physically stable).
Note that
$p$, unlike $H$, does not need to be bounded from below.
For a simple proof of the Pauli theorem, consider the operator
\begin{equation}\label{pauli1}
\hat{H}'\equiv e^{-i\epsilon\hat{T}/\hbar} \hat{H} 
e^{i\epsilon\hat{T}/\hbar} ,
\end{equation}
where $\epsilon$ is a positive parameter with the dimension of energy.
It is sufficient to consider the case of small $\epsilon$, so, 
by expanding the exponential functions and using (\ref{comrelTH}), 
one finds
\begin{equation}\label{pauli2}
\hat{H}' \approx \hat{H}-\epsilon .
\end{equation}
Now assume that the spectrum of $\hat{H}$ is bounded from below, 
i.e., that there exists a ground state $|\psi_0\rangle$ 
with the property $\hat{H}|\psi_0\rangle=E_0|\psi_0\rangle$, 
where $E_0$ is the minimal possible energy.
Consider the state
\begin{equation}\label{pauli3}
|\psi\rangle=e^{i\epsilon\hat{T}/\hbar}|\psi_0\rangle .
\end{equation}
Assuming that $\hat{T}$ is hermitian (i.e., that
$\hat{T}^{\dagger}=\hat{T}$) and using (\ref{pauli1}) and (\ref{pauli2}),
one finds
\begin{equation}\label{pauli4}
\langle\psi| \hat{H}| \psi\rangle=\langle\psi_0 |\hat{H}'| \psi_0\rangle
\approx E_0-\epsilon < E_0 .
\end{equation}
This shows that there exists a state $|\psi\rangle$ with the energy 
smaller than $E_0$. This is in contradiction with the asumption that 
$E_0$ is the minimal energy, which proves the theorem!  
There are attempts to modify some of the axioms of 
the standard form of quantum theory 
so that the commutation relation (\ref{comrelTH}) can be consistently
introduced (see, e.g., \cite{busch,bostr} and references therein), 
but the viability of such modified axioms of QM is not 
widely accepted among experts. 

Although (\ref{HtE}) is not a fundamental relation, in most practical 
situations it is still true that the uncertainty $\Delta E$ and the duration 
of the measurement process $\Delta t$ roughly satisfy the inequality 
(\ref{HtE}). However, there exists also an explicit counterexample 
that demonstrates that it is possible in principle to measure
energy with arbitrary accuracy during an arbitrarily short time-interval
\cite{ahar}.
It remains true that the characteristic evolution time of quantum states
is given by an uncertainty relation (\ref{HtE}), but this 
evolution time is not to be unequivocally identified with the duration 
of a quantum measurement. In this sense, the time-energy 
uncertainty relation (\ref{HtE}) is not equally fundamental as
the position-momentum uncertainty relation (\ref{Hxp}).

While different roles of space and time should not be surprising
in nonrelativistic QM, one may expect that space and time should play 
a more symmetrical role in relativistic QM. More on relativistic 
QM will be said in Sec.~\ref{RQM}, but here I only note that 
even in relativistic QM space and time do not play completely
symmetric roles, because even there integrals similar to (\ref{integ2})
have not a physical meaning, while those similar to (\ref{integ1})
have. Thus, even in relativistic QM, a time-energy uncertainty relation 
does not play a fundamental role.  

\section{QM implies that nature is fundamentally random} 

\subsection{Fundamental randomness as a myth}

QM is a theory that gives predictions on probabilities
for different outcomes of measurements. But this is not a 
privileged property of QM, classical statistical mechanics also does this.
Nevertheless, there is an important difference between QM and 
classical statistical mechanics. The latter is known to be an effective
approximative theory useful when not all fundamental
degrees of freedom are under experimental or theoretical control, 
while the underlying more fundamental classical dynamics
is completely deterministic. On the other hand, the usual form
of QM does not say anything about actual deterministic causes  
that lie behind the probabilistic quantum phenomena. 
This fact is often used to claim that QM implies that nature 
is fundamentally random. Of course, if the usual form of 
QM is really the ultimate truth, then it is true that nature 
is fundamentally random. But who says that the usual form of QM really 
{\em is} the ultimate truth? (A serious scientist will never 
claim that for any current theory.) {\it A priori}, one cannot exclude
the existence of some {\em hidden variables} (not described by the 
usual form of QM) that provide a deterministic cause for all 
seemingly random quantum phenomena. Indeed, from the experience 
with classical pseudorandom phenomena, the existence of such 
deterministic hidden variables seems a very natural hypothesis.
Nevertheless, QM is not that cheap; in QM there exist 
rigorous no-hidden-variable theorems. These theorems are often 
used to claim that hidden variables cannot exist and, consequently, 
that nature is fundamentally random. However, each theorem has 
assumptions. The main assumption is that hidden 
variables must reproduce the statistical predictions of QM.
Since these statistical predictions are verified experimentally, 
one is not allowed to relax this assumption. However, this assumption 
alone is not sufficient to provide a theorem. In the actual constructions
of these theorems, there are also some 
additional``auxiliary" assumptions, which, 
however, turn out to be physically crucial! Thus,
what these theorems actually prove, is that hidden variables, if exist, 
cannot have these additional assumed properties.
Since there is no independent proof that these additional 
assumed properties are necessary ingredients of nature, 
the assumptions of these theorems may not be valid.
(I shall discuss one version of these theorems in more detail in 
Sec.~\ref{NOREAL}.)
Therefore, the claim that QM implies fundamental randomness is 
a myth. 

\subsection{From analogy with classical statistical mechanics
to the Bohmian interpretation}

Some physicists, 
including one winner of the Nobel prize \cite{hooft},
very seriously take the possibility that some sort 
of deterministic hidden variables may underlie the usual form 
of QM.
In fact, the best known and most successful hidden-variable
extension of QM, the Bohmian interpretation, emerges rather 
naturally from the analogy with classical statistical mechanics.
To see this, consider a classical particle the position 
of which is not known with certainty. Instead, one deals 
with a statistical ensemble in which only the probability 
density $\rho({\bf x},t)$ is known. The probability must be 
conserved, i.e., $\int d^3x\, \rho=1$ for each $t$.
Therefore, the probability must satisfy the local conservation law
(known also as the continuity equation)
\begin{equation}\label{cons} 
\partial_t \rho +\nabla(\rho {\bf v})=0 ,
\end{equation}
where ${\bf v}({\bf x},t)$ is the velocity of the particle 
at the position ${\bf x}$ and the time $t$.
In the Hamilton-Jacobi formulation of classical 
mechanics, the velocity can be calculated as 
\begin{equation}\label{v}
{\bf v}({\bf x},t)=\frac{\nabla S({\bf x},t)}{m},
\end{equation}
where $S({\bf x},t)$ is a solution of the Hamilton-Jacobi equation
\begin{equation}\label{HJ}
\frac{(\nabla S)^2}{2m}+V({\bf x},t)=-\partial_t S ,
\end{equation}
$V({\bf x},t)$ is an arbitrary potential, and $m$ is the mass of the
particle.
The independent real equations (\ref{cons}) and (\ref{HJ}) can be
written in a more elegant form as a single complex equation.
For that purpose, one can introduce a complex function \cite{ros}
\begin{equation}\label{psi}
\psi=\sqrt{\rho} e^{iS/\hbar} ,
\end{equation}
where $\hbar$ is an arbitrary constant with the dimension of 
action, so that the exponent in (\ref{psi}) is dimensionless.
With this definition of $\psi$, Eqs. (\ref{cons}) and (\ref{HJ})
are equivalent to the equation
\begin{equation}\label{schcl}
\left( \frac{-\hbar^2 \nabla^2}{2m} +V-Q \right) \psi=
i\hbar \partial_t \psi ,
\end{equation}
where
\begin{equation}\label{Q}
Q\equiv -\frac{\hbar^2}{2m}\frac{\nabla^2 \sqrt{\rho}}{\sqrt{\rho}} .
\end{equation}
Indeed, by inserting (\ref{psi}) into (\ref{schcl}) and multiplying 
by $\psi^*$, 
it is straightforward to check that the real part of the resulting 
equation leads to (\ref{HJ}), while the imaginary part 
leads to (\ref{cons}) with (\ref{v}). 

The similarity of the 
classical equation (\ref{schcl}) with the quantum Schr\"odinger 
equation 
\begin{equation}\label{6.1}  
\left( \frac{-\hbar^2 \nabla^2}{2m} +V \right) \psi=
i\hbar \partial_t \psi            
\end{equation}
is obvious and remarkable! However, there are also some differences.
First, in the quantum case, the constant $\hbar$ is not arbitrary, 
but equal to the Planck constant divided by $2\pi$. The second
difference is the fact that (\ref{schcl}) contains 
the $Q$-term that is absent in the quantum case (\ref{6.1}).
Nevertheless, the physical interpretations are the same; in both cases, 
$|\psi({\bf x},t)|^2$ is the probability density of particle positions.
On the other hand, we know that classical mechanics is 
fundamentally deterministic.
This is incoded in the fact that Eq. (\ref{schcl}) alone does {\em not} 
provide a {\em complete} description of classical systems. 
Instead, one is also allowed to use Eq. (\ref{v}), which says
that {\em the velocity of the particle is determined whenever its
position is also determined}. The classical interpretation of 
this is that a particle {\em always} has a definite position 
and velocity and that the initial position and velocity uniquely 
determine the position and velocity at any time $t$. From this point 
of view, nothing seems more natural than to assume that 
an analogous statement is true also in the quantum case. 
This assumption represents the core of the Bohmian 
deterministic interpretation of QM.
To see the most obvious consequence of such a classical-like
interpretation of the Schr\"odinger equation,
note that the Schr\"odinger equation (\ref{6.1})
corresponds to a Hamilton-Jacobi equation in which 
$V$ in (\ref{HJ}) is replaced by $V+Q$. 
This is why $Q$ is often referred to as 
the {\em quantum potential}. The quantum potential induces a 
quantum force. Thus, a quantum particle trajectory satisfies  
a modified Newton equation
\begin{equation}\label{qnewt}
m\frac{d^2{\bf x}}{dt^2}=-\nabla(V+Q) .
\end{equation}
Such modified trajectories can be used to explain otherwise 
strange-looking quantum phenomena (see, e.g., \cite{pas}),
such as a two-slit experiment. 

Note that, so far, I have only discussed a single-particle
wave function $\psi({\bf x},t)$. When one generalizes this
to many-particle wave functions, an additional important feature
of the Bohmian interpretation becomes apparent -- the nonlocality. 
However, I delegate the explicit discussion of nonlocality to Sec.~\ref{L/NL}.

%Since $\sqrt{\rho}=|\psi|$, the quantity (\ref{Q}) explicitly depends 
%on $\psi$, which violates the linearity of (\ref{schcl}).
%In general, a linear equation is simpler than a nonlinear one.
%One could heuristically ``derive" quantum mechanics  
%from classical mechanics by modifying classical mechanics 
%such that $\psi$ satisfies a simpler, linear equation.
%The most obvious such modification
%by requiring that $\psi$ should satisfy a linear equation, which 
%motivates one to remove the $Q$-part  

\subsection{Random or deterministic?}

As we have seen above,
the analogy between classical statistical mechanics and 
QM can be used to interpret
QM in a deterministic manner. However, this analogy does not prove 
that such a deterministic interpretation of QM is correct.
Indeed, such deterministic quantum trajectories have never been 
directly observed. On the other hand, the Bohmian interpretation 
can explain why these trajectories are practically unobservable
\cite{duerr}, so the lack of experimental evidence does not 
disprove this interpretation. Most experts familiar with 
the Bohmian interpretation agree that the 
observable predictions of this 
interpretation are consistent with those of the standard 
interpretation, but they often prefer the standard interpretation 
because the standard interpretation seems simpler to them. 
This is because the standard interpretation of QM does not contain 
Eq. (\ref{v}). I call this {\em technical simplicity}. 
On the other hand, the advocates of the Bohmian interpretation 
argue that this technical extension 
of QM makes QM simpler on the {\em conceptual} level.  
Nevertheless, it seems that most contemporary physicists consider 
technical simplicity more important than conceptual simplicity, 
which explains why most physicists prefer the standard 
purely probabilistic interpretation of QM. In fact, 
by applying a QM-motivated technical criterion of simplicity, 
it can be argued 
that even classical statistical mechanics represented by (\ref{schcl})
can be considered complete, in which case even classical mechanics 
can be interpreted as a purely probabilistic theory \cite{nikolcl}.   
But the fact is that nobody knows with certainty whether the fundamental 
laws of nature are probabilistic or deterministic.

\section{QM implies that there is no reality besides the measured 
reality}
\label{NOREAL}

This is the central myth in QM and many other myths are based 
on this one. Therefore, it deserves a particularly careful 
analysis.

\subsection{QM as the ultimate scientific theory?}

On one hand, the claim that
``there is no reality besides the measured reality"
may seem to lie at the heart of the scientific method. 
All scientists agree that the empirical evidence is the 
ultimate criterion for acceptance or rejection of any 
scientific theory, so, from this point of view, such a claim
may seem rather natural. 
On the other hand, most scientists (apart from quantum physicists) 
do not find such a radical interpretation of the scientific method
appealing. In particular, many consider such an interpretation too 
antropomorfic (was there any reality before humans or living beings
existed?), while the history of science surprised us several times
by discovering that we (the human beings) are not so an important 
part of the universe as we thought we were. 
Some quantum physicists believe that QM is so 
deep and fundamental that it is not just a science that merely applies 
already prescribed scientific methods, but {\em the} science that 
answers the fundamental ontological and epistemological questions 
on the deepest possible level.
But is such a (certainly not modest) belief really founded? 
What are the true facts from which such a belief emerged? 
Let us see!  

\subsection{From a classical variable to a quantumlike representation}

Consider a simple real physical {\em classical}
variable $s$ that can attain only 
two different values, say $s_1=1$ and $s_2=-1$.
By assumption, such a variable cannot change continuously.
Neverheless, a quantity that can still change continuously
is the {\em probability} $p_n(t)$ that, at a 
given time $t$, the variable attains the value $s_n$.
The probabilities must satisfy
\begin{equation}\label{prob}
p_1(t) + p_2(t)=1 .
\end{equation}
The average value of $s$ is given by
\begin{equation}\label{av}
\langle s \rangle  = s_1p_1(t)+s_2p_2(t) .
\end{equation}
Although $s$ can attain only two values, $s_1=1$ and $s_2=-1$, 
the average value of $s$ can continuously change with time and attain 
an arbitrary value between $1$ and $-1$.  

The probabilities $p_n$ must be real and non-negative.
A simple formal way to provide this is to write
$p_n=\psi_n^*\psi_n$, where $\psi_n$ are 
auxiliary quantities that may be negative or even complex.      
It is also convenient to view the numbers $\psi_n$ (or $\psi_n^*$)  
as components of a {\em vector}. This vector can be represented either 
as a column
\begin{equation}\label{col}
|\psi\rangle \equiv \left( \begin{array}{c}
\psi_1 \\ \psi_2 \end{array} \right) ,
\end{equation}
or a row
\begin{equation}\label{row}
\langle \psi| \equiv (\psi_1^* , \psi_2^*) .
\end{equation}
The norm of this vector is
\begin{equation}\label{norm}
\langle \psi|\psi \rangle=\psi_1^*\psi_1+\psi_2^*\psi_2=p_1 + p_2 .
\end{equation}
Thus, the constraint (\ref{prob}) can be viewed as a constraint 
on the norm of the vector -- the norm must be unit.
By introducing two special unit vectors
\begin{equation}\label{basis}
|\phi_1\rangle = |\!\uparrow\,\rangle \equiv 
\left( \begin{array}{c}
1 \\ 0 \end{array} \right) , \;\;\;\;
|\phi_2\rangle = |\!\downarrow\,\rangle \equiv \left( \begin{array}{c}
0 \\ 1 \end{array} \right) ,
\end{equation}  
one finds that the probabilities can be expressed in terms of 
vector products as
\begin{equation}\label{p12}
p_1=|\langle \phi_1|\psi\rangle|^2 , \;\;\;\;
p_2=|\langle \phi_2|\psi\rangle|^2 .
\end{equation}
It is also convenient to introduce a diagonal {\em matrix} $\sigma$ that 
has the values $s_n$ at the diagonal and the zeros at all other places:
\begin{equation}\label{sigma}
\sigma\equiv \left( \begin{array}{cc}
s_1 & 0 \\
0 & s_2 
\end{array} \right)
= \left( \begin{array}{cc}
1 & 0 \\
0 & -1
\end{array} \right) .
\end{equation}   
The special vectors (\ref{basis}) have the property 
\begin{equation}\label{s12}
\sigma |\phi_1\rangle=s_1|\phi_1\rangle , \;\;\;\;
\sigma |\phi_2\rangle=s_2|\phi_2\rangle ,
\end{equation}
which shows that (i) the vectors (\ref{basis}) are the eigenvectors of 
the matrix $\sigma$ and (ii) the eigenvalues of $\sigma$ are 
the allowed values $s_1$ and $s_2$.  
The average value (\ref{av}) can then be formally written as
\begin{equation}
\langle s \rangle =\langle \psi(t)|\sigma |\psi(t) \rangle .
\end{equation}

What has all this to do with QM? First, a discrete spectrum 
of the allowed values is typical of quantum systems; after all, 
discrete spectra are often referred to as ``quantized" spectra,
which, indeed, is why QM attained its name.
(Note, however, that it would be misleading
to claim that quantized spectra is the most fundamental property of
quantum systems. Some quantum variables, such as the position of a particle,
do not have quantized spectra.)
A discrete spectrum contradicts some common
prejudices on classical physical systems because
such a spectrum does not allow a continuous change of the variable.
Nevertheless, a discrete spectrum alone does not yet imply
quantum physics. The formal representation of probabilities and average 
values in terms of complex numbers, vectors, and matrices
as above is, of course, inspired by the 
formalism widely used in QM; yet, this representation by itself 
does not yet imply QM. The formal representation in terms of
complex numbers, vectors, and matrices can still be interpreted 
in a classical manner.

\subsection{From the quantumlike representation to quantum variables}

The really interesting things that deviate significantly from the  
classical picture emerge when one recalls the 
following formal algebraic properties of vector spaces:
Consider an arbitrary $2\times 2$ unitary matrix $U$, $U^{\dagger}U=1$.
(In particular, $U$ may or may not be time dependent.) 
Consider a formal transformation 
\begin{eqnarray}    
& |\psi'\rangle =U|\psi\rangle , \;\;\;\;
\langle\psi'|=\langle\psi|U^{\dagger} , & \nonumber \\ 
& \sigma'=U\sigma U^{\dagger} . &
\end{eqnarray}
(This transformation refers to {\em all} vectors $|\psi\rangle$
or $\langle\psi|$, 
including the eigenvectors $|\phi_1\rangle$ and $|\phi_2\rangle$.)  
In the theory of vector spaces, such a transformation can be interpreted 
as a new representation of the {\em same} vectors. 
Indeed, such a transformation does not change the physical properties, 
such as the norm of the vector (\ref{norm}), the probabilities
(\ref{p12}), and the eigenvalues in (\ref{s12}), calculated 
in terms of the primed quantities $|\psi'\rangle$, $\langle\psi'|$, 
and $\sigma'$. This means that {\em the explicit representations}, 
such as those in (\ref{col}), (\ref{row}), (\ref{basis}), and (\ref{sigma}),
{\em are irrelevant}. Instead, {\em the only physically relevant
properties are abstract, representation-independent quantities, 
such as scalar products and the spectrum of eigenvalues}.
What does it mean physically? One possibility is not to take it 
too seriously, as it is merely an artefact of an artificial 
vector-space representation of certain physical quantities.
However, the history of theoretical physics teaches us that formal 
mathematical symmetries often have a deeper physical message.
So let us try to take it seriously, to see where it 
will lead us. Since the representation is not relevant, 
it is natural to ask if there are {\em other} matrices
(apart from $\sigma$) that do {\em not} have the form 
of (\ref{sigma}), but still have the same spectrum of eigenvalues
as $\sigma$? The answer is {\em yes}! 
But then we are in a very strange, if not paradoxical, position; 
we have started with a consideration of a {\em single} physical variable 
$s$ and arrived at a result that seems to suggest the existence 
of some {\em other}, equally physical, variables. 
As we shall see, this strange result lies at the heart of the 
(also strange) claim
that there is no reality besides the measured reality.  
But let us not jump to the conclusion too early! Instead, let us 
first study the mathematical properties of these 
additional physical variables.

Since an arbitrary $2\times 2$ matrix is defined by 4 independent 
numbers, each such matrix can be written as a linear combination 
of 4 independent matrices. One convenient choice of 4 independent 
matrices is
\begin{eqnarray}\label{pauli}    
&
1 \equiv \left( \begin{array}{cc}
1 & 0 \\
0 & 1
\end{array} \right) , \;\;\;\;
\sigma_1 \equiv \left( \begin{array}{cc}
0 & 1 \\
1 & 0
\end{array} \right) , & \nonumber \\
&
\sigma_2 \equiv \left( \begin{array}{cc}
0 & -i \\
i & 0
\end{array} \right) , \;\;\;\;
\sigma_3 \equiv \left( \begin{array}{cc}
1 & 0 \\
0 & -1
\end{array} \right) . &
\end{eqnarray}
The matrix $\sigma_3$ is nothing but a renamed matrix $\sigma$ in
(\ref{sigma}). The matrices $\sigma_i$, known also 
as Pauli matrices, are chosen so that they satisfy 
the familiar symmetrically looking commutation relations 
\begin{equation}\label{comsig}
[\sigma_j,\sigma_k]=2i\epsilon_{jkl}\sigma_l  
\end{equation}
(where summation over repeated indices is understood).
The matrices $\sigma_i$ are all hermitian, 
$\sigma_i^{\dagger}=\sigma_i$, which implies that their eigenvalues 
are real. Moreover, all three $\sigma_i$ have the eigenvalues 
$1$ and $-1$. The most explicit way to see this is to construct 
the corresponding eigenvectors. The eigenvectors of 
$\sigma_3$ are $|\!\uparrow_3\rangle \equiv |\!\uparrow\,\rangle$ 
and $|\!\downarrow_3\rangle \equiv |\!\downarrow\,\rangle$
defined in (\ref{basis}), with the eigenvalues $1$ and $-1$, respectively.
Analogously, it is easy to check that the eigenvectors of $\sigma_1$ are
\begin{eqnarray}\label{basis1}
& |\!\uparrow_1\rangle=\frac{1}{\sqrt{2}}
\left( \begin{array}{c} 1 \\ 
                        1 \end{array} \right)=
\displaystyle\frac{ 
|\!\uparrow\,\rangle + |\!\downarrow\,\rangle }{\sqrt{2}} , &
\nonumber \\
& |\!\downarrow_1\rangle=\frac{1}{\sqrt{2}}
\left( \begin{array}{c} 1 \\ 
                        -1 \end{array} \right)=
\displaystyle\frac{
|\!\uparrow\,\rangle - |\!\downarrow\,\rangle }{\sqrt{2}} , &
\end{eqnarray}
with the eigenvalues $1$ and $-1$, respectively, while the 
eigenvectors of $\sigma_2$ are
\begin{eqnarray}\label{basis2}
& |\!\uparrow_2\rangle=\frac{1}{\sqrt{2}}
\left( \begin{array}{c} 1 \\ 
                        i \end{array} \right)=
\displaystyle\frac{
|\!\uparrow\,\rangle + i|\!\downarrow\,\rangle }{\sqrt{2}} , &
\nonumber \\
& |\!\downarrow_2\rangle=\frac{1}{\sqrt{2}}
\left( \begin{array}{c} 1 \\
                        -i \end{array} \right)=
\displaystyle\frac{
|\!\uparrow\,\rangle - i|\!\downarrow\,\rangle }{\sqrt{2}} , &  
\end{eqnarray}
with the same eigenvalues $1$ and $-1$, respectively. 
 
The commutation relations (\ref{comsig})
are invariant under the unitary 
transformations $\sigma_i\rightarrow \sigma_i'=U\sigma_i U^{\dagger}$.
This suggests that the commutation relations themselves are more 
physical than the explicit representation given by (\ref{pauli}).
Indeed, the commutation relations (\ref{comsig}) can be recognized 
as the algebra of the generators of the group of rotations in 3 
spacial dimensions. There is nothing quantum mechanical about that;
in classical physics, matrices represent {\em operators}, that is, 
abstract objects that act on vectors by changing (in this case, rotating) 
them. However, in the usual formulation of classical physics, 
there is a clear distinction between
operators and physical variables -- the latter are not represented 
by matrices. In contrast, in our formulation, the matrices $\sigma_i$ have a 
double role; mathematically, they are operators (because they act on 
vectors $|\psi\rangle$), while physically, they represent physical 
variables. From symmetry, it is natural to assume that all three $\sigma_i$
variables are equally physical. This assumption is one of the central
assumptions of QM that makes it different from classical mechanics.
For example, the spin operator of spin $\frac{1}{2}$ particles in QM
is given by
\begin{equation}
S_i=\frac{\hbar}{2}\sigma_i ,
\end{equation}
where the 3 labels $i=1,2,3$ correspond to 3 space directions
$x,y,z$, respectively. Thus, in the case of spin,
it is clear that $\sigma_3\equiv \sigma_z$ cannot be 
more physical than $\sigma_1\equiv \sigma_x$ or 
$\sigma_2\equiv \sigma_y$, despite the fact 
that $\sigma_3$ corresponds to the initial physical variable 
with which we started our considerations. On the other hand, 
the fact that the new variables $\sigma_1$ and $\sigma_2$ 
emerged from the initial variable $\sigma_3$ suggests that,
in some sense, these 
3 variables are not really completely independent. Indeed, 
a nontrivial relation among them is incoded in the nontrivial 
commutation relations (\ref{comsig}). In QM, two variables 
are really independent only if their commutator vanishes. 
(For example, recall that, unlike (\ref{comsig}), 
the position operators $x_i$ and 
the momentum operators $p_i$ in QM satisfy 
$[x_i,x_j]=[p_i,p_j]=0$. In fact, this is the ultimate reason why 
the most peculiar aspects of QM are usually discussed 
on the example of spin variables, rather than on position or 
momentum variables.)

\subsection{From quantum variables to quantum measurements}
  
Now consider the state
\begin{equation}\label{50:50}
|\psi\rangle = \frac{
|\!\uparrow\,\rangle + |\!\downarrow\,\rangle }{\sqrt{2}} .
\end{equation}
In our initial picture, this state merely represents a situation 
in which there are $50:50$ chances that the system has the value of 
$s$ equal to either $s=1$ or $s=-1$. Indeed, 
if one performs a measurement to find out what that value 
is, one will obtain one and only one of these two values. 
By doing such a measurement, the observer gaines new information about 
the system. For example, if the value turns out to be $s=1$, this 
gain of information can be described by a ``collapse"
\begin{equation}\label{coll}
|\psi\rangle \rightarrow |\!\uparrow\,\rangle ,
\end{equation}
as the state $|\!\uparrow\,\rangle$ corresponds to a situation in which 
one is certain that $s=1$.
At this level, there is nothing mysterious and nothing intrinsically
quantum about this collapse. However, in QM, the state (\ref{50:50})
contains more information than said above! 
(Otherwise, there would be no 
physical difference between the two different states in   
(\ref{basis1}).) From (\ref{basis1}), we see that {\em the ``uncertain" state
(\ref{50:50}) corresponds to a situation in which one is absolutely certain 
that the value of the variable $\sigma_1$ is equal to $1$}. On the other
hand, if one performs the measurement of $s=\sigma_3$ and obtains 
the value as in (\ref{coll}), then 
the postmeasurement state
\begin{equation}\label{50:50.2}
|\!\uparrow\,\rangle = \displaystyle\frac{
|\!\uparrow_1\rangle + |\!\downarrow_1\rangle }{\sqrt{2}}
\end{equation}
implies that the value of $\sigma_1$ is no longer known with certainty.
This means that, in some way, the measurement of $\sigma_3$ destroys 
the {\em information} on $\sigma_1$. But the crucial question is
not whether the information on $\sigma_1$ has been destroyed, 
but rather whether the {\em value itself} of $\sigma_1$ has been destroyed. 
In other words, is it possible that all the time, 
irrespective of the performed measurements, $\sigma_1$ has the value $1$?
The fact is that if this were the case, then it would contradict 
the predictions of QM! The simplest way to see this is to observe 
that, after the first measurement with the result (\ref{coll}), 
one can perform a new measurement, in which one measures $\sigma_1$.
From (\ref{50:50.2}), one sees that there are $50\%$ chances 
that the result of the new measurement will give the value $-1$.
That is, there is $0.5\cdot 0.5=0.25$ probability that the sequence of the two 
measurements will correspond to the collapses
\begin{equation}\label{2coll}
|\!\uparrow_1\rangle \rightarrow |\!\uparrow\,\rangle \rightarrow
|\!\downarrow_1\rangle .
\end{equation}
In (\ref{2coll}), the initial value of $\sigma_1$ is $1$, while the 
final value of $\sigma_1$ is $-1$. Thus, QM predicts that the value of 
$\sigma_1$ may change during the process of the two measurements.
Since the predictions of QM are in agreement with experiments, 
we are forced to accept this as a fact. This demonstrates that 
QM is {\em contextual}, that is, that the measured values depend on the 
context, i.e., on the measurement itself. This property 
by itself is still not intrinsically quantum, in classical physics 
the result of a measurement may also depend on the measurement.
Indeed, in classical mechanics there is nothing mysterious about it;
there, a measurement is a physical process that, as any other physical 
process, may influence the values of the physical variables. But 
can we talk about the value of the variable irrespective of 
measurements? From a purely experimental point of view, we certainly cannot.
Here, however, we are talking about theoretical physics. 
So does the theory allow to talk about that? Classical theory 
certainly does. But what about QM? If {\em all} theoretical knowledge 
about the system is described by the state $|\psi\rangle$, 
then quantum theory does {\em not} allow that! This is the fact. 
But, can we be sure that we shall never discover some more complete theory 
than the current form of QM, so that this more complete theory 
will talk about theoretical values of variables irrespective of 
measurements? From the example above, it is not possible to
draw such a conclusion. Nevertheless, physicists are trying 
to construct more clever examples from which such a conclusion 
could be drawn. Such examples are usually referred to as 
``no-hidden-variable theorems". But what do these theorems 
really prove? Let us see!     

\subsection{From quantum measurements to no-hidden-variable 
theorems}

To find such an example, consider a system consisting of 
{\em two independent} subsystems, such that each subsystem
is characterized by a variable that can attain 
only two values, $1$ and $-1$. The word ``independent" 
(which will turn out to be the crucial word) means that the corresponding
operators commute and that the Hamiltonian does not contain 
an interaction term between these two variables. 
For example, this can be a system with two free particles, 
each having spin $\frac{1}{2}$. In this case, the state 
$|\!\uparrow\,\rangle \otimes |\!\downarrow\,\rangle
\equiv |\!\uparrow\,\rangle |\!\downarrow\,\rangle$
corresponds to the state in which the first particle is in the 
state $|\!\uparrow\,\rangle$, while the second particle 
is in the state $|\!\downarrow\,\rangle$. 
(The commutativity of the corresponding variables is 
provided by the operators $\sigma_j\otimes 1$ and 
$1\otimes\sigma_k$ that correspond to the variables of the 
first and the second subsystem, respectively.)   
Instead of (\ref{50:50}), consider the state 
\begin{equation}\label{EPR}
|\psi\rangle = \frac{
|\!\uparrow\,\rangle |\!\downarrow\,\rangle + 
|\!\downarrow\,\rangle |\!\uparrow\,\rangle }{\sqrt{2}} .
\end{equation}
This state constitutes the basis for the famous   
Einstein-Podolsky-Rosen-Bell paradox. This state says that 
if the first particle is found in the state $|\!\uparrow\,\rangle$,
then the second particle will be found in the state
$|\!\downarrow\,\rangle$, and vice versa. In other words,   
the second particle will always take a direction opposite
to that of the first particle. In an oversimplified version of 
the paradox, one can wonder how the second particle 
knows about the state of the first particle, given the 
assumption that there is no interaction between the two 
particles? However, this oversimplified version of the 
paradox can be easily resolved in classical terms by observing 
that the particles do not necessarily need to interact, because
they could have their (mutually opposite) values all the time 
even before the measurement, while the only role of the measurement 
was to reveal these values. The case of a single particle 
discussed through Eqs. (\ref{50:50})-(\ref{2coll}) suggests 
that a true paradox can only be obtained when one assumes 
that the variable corresponding to $\sigma_1$ or $\sigma_2$ 
is also a physical variable. Indeed, this is what has been obtained 
by Bell \cite{bell}. The paradox can be expressed in terms of 
an {\em inequality} that the correlation functions among different 
variables must obey if the measurement is merely a revealation 
of the values that the noninteracting particles had before 
the measurement. (For more detailed pedagogic expositions, 
see \cite{merm1,laloe}.)       
The predictions of QM turn out to be in contradiction with this 
Bell inequality. The experiments violate the Bell inequality
and confirm the predictions of QM (see \cite{genov} for a recent review).
This is the fact! However, 
instead of presenting a detailed derivation 
of the Bell inequality, for pedagogical purposes 
I shall present a simpler example 
that does not involve inequalities, but leads to the same
physical implications.

The first no-hidden-variable theorem without inequalities has 
been found by Greenberger, Horne, and Zeilinger \cite{GHZ}
(for pedagogic expositions, see \cite{merm2,jord,laloe}), for a system 
with 3 particles. However, the simplest such theorem is the one discovered
by Hardy \cite{hardy} (see also \cite{hardy-more}) that, 
like the one of Bell, involves only 2 particles. Although pedagogic
expositions of the Hardy result also exist \cite{jord,merm3,laloe},
since it still seems not to be widely known in the physics community,
here I present a very simple exposition of the Hardy result (so simple 
that one can really wonder why the Hardy result was not discovered 
earlier).    
Instead of (\ref{EPR}), consider a slightly more complicated state
\begin{equation}\label{H1}
|\psi\rangle = \frac{
|\!\downarrow\,\rangle |\!\downarrow\,\rangle +
|\!\uparrow\,\rangle |\!\downarrow\,\rangle +
|\!\downarrow\,\rangle |\!\uparrow\,\rangle }{\sqrt{3}} .
\end{equation}
Using (\ref{basis1}), we see that this state can also be written in two 
alternative forms as
\begin{equation}\label{H2}
|\psi\rangle = \frac{
\sqrt{2} |\!\downarrow\,\rangle |\!\uparrow_1\,\rangle +
|\!\uparrow\,\rangle |\!\downarrow\,\rangle }{\sqrt{3}} ,
\end{equation}
\begin{equation}\label{H3}
|\psi\rangle = \frac{
\sqrt{2} |\!\uparrow_1\rangle |\!\downarrow\,\rangle +
|\!\downarrow\,\rangle |\!\uparrow\,\rangle }{\sqrt{3}} .
\end{equation} 
From these 3 forms of $|\psi\rangle$, we can infer the following: \newline
(i) From (\ref{H1}), at least one of the particles is in the state 
$|\!\downarrow\,\rangle$. \newline
(ii) From (\ref{H2}), if the first particle is in the state 
$|\!\downarrow\,\rangle$, then the second particle is in the state 
$|\!\uparrow_1\,\rangle$. \newline
(iii) From (\ref{H3}), if the second particle is in the state
$|\!\downarrow\,\rangle$, then the first particle is in the state
$|\!\uparrow_1\,\rangle$. \newline
Now, by classical reasoning, from (i), (ii), and (iii) one infers that
\newline
(iv) It is impossible that both particles are in the state 
$|\!\downarrow_1\,\rangle$. \newline
But is (iv) consistent with QM? If it is, then 
$\langle \downarrow_1\!|\langle \downarrow_1\!|\psi\rangle$ must 
be zero. However, using (\ref{H2}), 
$\langle \downarrow_1\!|\!\uparrow_1\rangle =0$, and
an immediate consequence of (\ref{basis1}) 
$\langle \downarrow_1\!|\!\uparrow\rangle = 
-\langle \downarrow_1\!|\!\downarrow\rangle =1/\sqrt{2}$,
we see that
\begin{equation}
\langle \downarrow_1\!|\langle \downarrow_1\!|\psi\rangle =
\frac{
\langle \downarrow_1\!|\!\uparrow\,\rangle
\langle \downarrow_1\!|\!\downarrow\,\rangle }{\sqrt{3}}
=\frac{-1}{2\sqrt{3}} ,
\end{equation} 
which is not zero. Therefore, (iv) is wrong in QM; there is a finite 
probability for both particles to be in the state
$|\!\downarrow_1\rangle$. This is the fact!
But what exactly is wrong with the reasoning that led to (iv)?
The fact is that there are {\em several}(!) possibilities. Let us 
briefly discuss them.

\subsection{From no-hidden-variable theorems to physical interpretations}

One possibility is that classical logic cannot be used 
in QM. Indeed, this motivated the development of a branch of QM
called {\em quantum logic}. However, most physicists 
(as well as mathematicians) consider 
a deviation from classical logic too radical. 

Another possibility is that only one matrix, say $\sigma_3$, 
corresponds to a genuine physical variable. In this case, the 
true state of a particle can be $|\!\uparrow\,\rangle$ or 
$|\!\downarrow\,\rangle$, but not a state such as
$|\!\uparrow_1\rangle$ or $|\!\downarrow_1\rangle$. Indeed, such a 
possibility corresponds to our starting picture in which there 
is only one physical variable called $s$ that was later artifically 
represented by the matrix $\sigma\equiv\sigma_3$. Such an 
interpretation may seem reasonable, at least for some physical 
variables. However, if $\sigma_3$ corresponds to the spin
in the $z$-direction, then it does not seem reasonable that 
the spin in the $z$-direction is more physical than that in 
the $x$-direction or the $y$-direction. Picking up one 
preferred variable breaks the symmetry, which, at least in some cases, 
does not seem reasonable.

The third possibility is that one should distinguish between 
the claims that ``the system {\em has} a definite value of a variable" 
and ``the system is {\em measured} to have a definite value of a variable".
This interpretation of QM is widely accepted. According to this 
interpretation, the claims (i)-(iii) refer only to the results of 
measurements. These claims assume that $\sigma=\sigma_3$ is measured 
for at least one of the particles. Consequently, the claim 
(iv) is valid only if this assumption is fulfilled. In contrast, 
if $\sigma_3$ is not measured at all, then it is possible to measure 
both particles to be in the state $|\!\downarrow_1\rangle$.
Thus, the paradox that (iv) seems to be both correct and incorrect 
is merely a manifestation of quantum contextuality. 
In fact, {\em all} no-hidden-variable theorems can be viewed 
as manifestations of quantum contextuality. 
However, there are at least two drastically different versions of this
quantum-contextuality interpretation. 
In the first version, it does not make sense even to talk about 
the values that are not measured. I refer to this version 
as the {\em orthodox} interpretation of QM. (The orthodox 
interpretation can be further divided into a {\em hard} version 
in which it is claimed that such unmeasured values simply
do not exist, and a {\em soft} version according to which
such values perhaps might exist, but  
one should not talk about them because one cannot know about the 
existence of something that is not measured.)
In the second version, the variables have some values even 
when they are not measured, but the process of measurement is a 
physical process that may influence these values. The second 
version assumes that the standard formalism of QM is not complete, 
i.e., that even a more accurate description of physical systems 
is possible than that provided by standard QM. According to this version, 
``no-hidden-variable" theorems (such as the one of Bell or Hardy)
do not really prove that hidden variables cannot exist, because
these theorems {\em assume} that there are no interactions 
between particles, while this assumption may be violated  
at the level of hidden variables. 
The most frequent argument for the validity of this assumption
is the locality principle, which then must be violated 
by any hidden-variable completion of standard QM.
However, since the assumption of the absence of interaction 
between particles is much more general than the assumption
that it is the locality principle that forbids 
such an interaction, and, at the same time, since the 
the discussion of the locality principle deserves 
a separate section, I delegate the detailed discussion of 
locality to the next section, Sec.~\ref{L/NL}. 

Most pragmatic physicists seem to (often tacitly) accept the soft-orthodox 
interpretation. From a pragmatic point of view, such an 
attitude seems rather reasonable. However, physicists 
who want to understand QM at the deepest possible level can hardly 
be satisfied with the soft version of the orthodox interpretation.
They are forced either to adopt the hard-orthodox interpretation 
or to think about the alternatives (like hidden 
variables, preferred variables, or quantum logic). 
Among these physicists that cope with the foundations of QM
at the deepest level, 
the hard-orthodox point of view seems to dominate.
(If it did not dominate, then I would not call it ``orthodox").
However, even the advocates of the hard-orthodox interpretation 
do not really agree what exactly this interpretation means.
Instead, there is a number of subvariants of the
hard-orthodox interpretation that differ
in the fundamental ontology of nature.
Some of them are rather antropomorphic, by attributing 
a fundamental role to the observers.
However, most of them attempt to avoid antropomorphic 
ontology, for example by proposing that  
the concept of {\em information} on reality is more fundamental 
than the concept of reality itself \cite{zeil}, 
or that reality is relative or ``relational" 
\cite{rov1,rov2}, or that correlations among variables exist, 
while the variables themselves do not \cite{mermcor}.
Needless to say, all such versions of the hard-orthodox
interpretation necessarily involve deep (and dubious)
philosophical assumptions and postulates.
To avoid philosophy, an alternative is to adopt 
a softer version of the orthodox interpretation 
(see, e.g., \cite{medina}). The weakness of the 
soft versions is the fact that they do not even try to 
answer fundamental questions one may ask, but 
their advocates often argue that these questions are not 
physical, but rather metaphysical or philosophical.    

Let us also discuss in more detail the possibility that one variable 
is more physical than the others, that only this preferred variable 
corresponds to the genuine physical reality. Of course, 
it does not seem reasonable that spin in the 
$z$-direction is more physical than that in the $x$- or the $y$-direction. 
However, it is not so unreasonable that, for example, 
the particle position is a more fundamental variable 
than the particle momentum or energy. (After all, 
most physicists will agree that this is so 
in classical mechanics, despite the fact that the Hamiltonian 
formulation of classical mechanics treats position and momentum 
on an equal footing.) Indeed, in practice, all quantum measurements 
eventually reduce to an observation of the {\em position} of 
something (such as the needle of the measuring apparatus). 
In particular, the spin of a particle is measured by a 
Stern-Gerlach apparatus, in which the magnetic field causes 
particles with one orientation of the spin to change their 
direction of motion to one side, and those with the opposite 
direction to the other. Thus, one does not really observe 
the spin itself, but rather the position of the particle.
In general, assume that one wants to measure the value of the 
variable described by the operator $A$. It is convenient 
to introduce an orthonormal basis $\{ |\psi_a\rangle \}$
such that each $|\psi_a\rangle$ is an eigenvector of the 
operator $A$ with the eigenvalue $a$.
The quantum state can be expanded in this basis as
\begin{equation}
|\psi\rangle = \sum_a c_a |\psi_a\rangle ,
\end{equation}
where (assuming that the spectrum of $A$ is not degenerate)
$|c_a|^2$ is the probability that the variable 
will be measured to have the value $a$. To perform a measurement, 
one must introduce the degrees of freedom of the measuring apparatus, 
which, before the measurement, is described by some state 
$|\phi\rangle$. In an ideal measurement, the interaction between
the measured degrees of freedom and the degrees of 
freedom of the measuring apparatus must be such that the total quantum
state exhibits entanglement between these two degrees of freedom, 
so that the total state takes the form
\begin{equation}\label{meas}
|\Psi\rangle = \sum_a c_a |\psi_a\rangle |\phi_a\rangle ,
\end{equation} 
where $|\phi_a\rangle$ are orthonormal states of the measuring apparatus.
Thus, whenever the measuring apparatus is 
found in the state $|\phi_a\rangle$, one can be certain 
(at least theoretically) that the state 
of the measured degree of freedom is given by $|\psi_a\rangle$. 
Moreover, from (\ref{meas}) it is clear that the probability for 
this to happen is equal to $|c_a|^2$, the same probability   
as that without introducing the measuring apparatus.
Although the description of the quantum measurement as described 
above is usually not discussed in practical textbooks on QM, 
it is actually a part of the standard form of quantum theory
and does not depend on the interpretation. (For modern practical 
introductory lectures on QM in which the theory of measurement
is included, see, e.g., \cite{lectqm}.)  
What this theory of quantum measurement suggests is that, 
in order to reproduce the statistical predictions of standard QM, it is not 
really necessary that all hermitian operators called ``observables" 
correspond to genuine physical variables. Instead, it is 
sufficient that only one or a few preferred variables that are really 
measured in practice correspond to genuine physical variables, 
while the rest of the ``observables" are merely 
hermitian operators that do not correspond to true 
physical reality \cite{naive}. This is actually the reason 
why the Bohmian interpretation discussed in the preceding section,
in which the preferred variables are the particle positions, is able 
to reproduce the quantum predictions on {\em all} 
quantum observables, such as momentum, energy, spin, etc.
Thus, the Bohmian interpretation combines two possibilities discussed 
above: one is the existence of the preferred variable (the particle position) 
and the other is the hidden variable (the particle position existing even 
when it is not measured).

To conclude this section, QM does not prove that there is no 
reality besides the measured reality. Instead, there are several 
alternatives to it. In particular, such reality may exist, 
but then it must be contextual (i.e., must depend 
on the measurement itself.) The simplest (although not necessary) 
way to introduce such reality is to postulate it only for one 
or a few preferred quantum observables.    
  
\section{QM is local/nonlocal}
\label{L/NL}

\subsection{Formal locality of QM}

Classical mechanics is local. This means that a physical quantity 
at some position ${\bf x}$ and time $t$
may be influenced by another physical 
quantity only if this other physical quantity is attached to the 
same ${\bf x}$ and $t$. For example, two spacially separated local objects 
cannot communicate directly, but only via a third physical object 
that can move from one object to the other. In the case of $n$ particles,
the requirement of locality can be written as a requirement 
that the Hamiltonian 
$H({\bf x}_1, \ldots,{\bf x}_n,{\bf p}_1, \ldots,{\bf p}_n)$ 
should have the form
\begin{equation}\label{hcl}
H=\sum_{l=1}^{n} H_l({\bf x}_l,{\bf p}_l) .
\end{equation}
In particular, a nontrivial 2-particle potential of the form 
$V({\bf x}_1-{\bf x}_2)$ is forbidden by the principle of 
locality. Note that such a potential is {\em not} forbidden
in Newtonian classical mechanics. However, known fundamental 
interactions are relativistic interactions that do not allow 
such instantaneous communications.
At best, such a nonlocal potential can be used as an approximation
valid when the particles are sufficiently close to each other
and their velocities are sufficiently small.

The quantum Hamiltonian is obtained from the corresponding classical
Hamiltonian by a replacement of classical positions and momenta 
by the corresponding quantum operators. Thus, the quantum Hamiltonian 
takes the same local form as the classical one.
Since the Schr\"odinger equation 
\begin{equation}\label{schloc}
H|\psi(t)\rangle =i\hbar\partial_t|\psi(t)\rangle 
\end{equation}
is based on this local Hamiltonian, any change of the wave function 
induced by the Schr\"odinger equation (\ref{schloc}) 
is local. This is the fact. For this reason, it is often claimed 
that QM is local to the same extent as classical mechanics is. 

\subsection{(Non)locality and hidden variables}

The principle of locality is often used as the crucial argument against 
hidden variables in QM. For example, consider two particles entangled 
such that their wave function 
(with the spacial and temporal dependence of wave functions suppressed)
takes the form 
(\ref{H1}). Such a form of the wave function can be kept even 
when the particles become spacially separated. As we have seen, 
the fact that (iv) is inconsistent with QM can be interpreted 
as QM contextuality. However, we have seen that
there are two versions of QM contextuality --
the orthodox one and the hidden-variable one.
The principle of locality excludes the hidden-variable version 
of QM contextuality, because this version 
requires interactions between the two particles, which are 
impossible when the particles are (sufficiently) spacially separated.
However, it is important to emphasize that the principle of 
locality is an {\em assumption}. We know that the Schr\"odinger equation 
satisfies this principle, but we do not know if this principle 
must be valid for {\em any} physical theory. In particular, 
subquantum hidden variables might not satisfy this principle.
Physicists often object that nonlocal interactions contradict
the theory of relativity. However, there are several responses
to such objections. First, the theory of relativity is just as 
any other theory -- nobody can be certain that this theory is absolutely 
correct at all (including the unexplored ones) levels. 
Second, nonlocality by itself does not necessarily 
contradict relativity. For example, a local 
relativistic-covariant field theory (see Sec.~\ref{QFT})
can be defined by an action 
of the form $\int d^4x {\cal L}(x)$, where ${\cal L}(x)$ is the local 
Lagrangian density transforming (under arbitrary
coordinate transformations) as a scalar density. A nonlocal 
action may have a form $\int d^4x \int d^4x'{\cal L}(x,x')$. If 
${\cal L}(x,x')$ transforms as a bi-scalar density, then such a nonlocal 
action is relativistically covariant. Third, the nonlocality needed 
to explain quantum contextuality requires {\em instantaneous} 
communication, which is often claimed to be excluded by the theory 
of relativity, as the velocity of light is the maximal possible 
velocity allowed by the theory of relativity. However, this is 
actually a myth in the theory of relativity; this theory by itself 
does {\em not} exclude faster-than-light communication. 
It excludes it {\em only} if some {\em additional} assumptions 
on the nature of matter are used. The best known counterexample 
are {\em tachyons} \cite{tachyon} -- hypothetical particles
with negative mass squared 
that move faster than light and fully respect the theory of relativity.
Some physicists argue that faster-than-light communication 
contradicts the principle of causality, but this is also 
nothing but a myth \cite{liberati,nikolcaus}. (As shown in 
\cite{nikolcaus}, this myth can be traced back to one of the 
most fundamental myths in physics according to which 
time fundamentally differs from space by having a property
of ``lapsing".)           
Finally, some physicists find absurd or difficult even to conceive 
physical laws in which information between distant objects is 
transferred instantaneously. It is ironic that they probably 
had not such mental problems many years ago when they did not know 
about the theory of relativity but {\em did} know about the 
Newton instantaneous law of gravitation or the Coulomb 
instantaneous law of electrostatics.  
To conclude this paragraph, hidden variables, if exist, must 
violate the principle of locality, which may or may not 
violate the theory of relativity. 

To illustrate nonlocality of hidden variables, 
I consider the example of the Bohmian interpretation.
For a many-particle wave function 
$\Psi({\bf x}_1,\ldots,{\bf x}_n,t)$ that describes $n$ 
particles with the mass $m$, it is 
straightforward to show that the generalization of (\ref{Q}) is
\begin{equation}\label{Qn}
Q({\bf x}_1,\ldots,{\bf x}_n,t)
= -\frac{\hbar^2}{2m} 
\displaystyle\frac{\displaystyle\sum_{l=1}^{n}
\nabla^2_l \sqrt{\rho({\bf x}_1,\ldots,{\bf x}_n,t)}}
{\sqrt{\rho({\bf x}_1,\ldots,{\bf x}_n,t)}} .
\end{equation}
When the wave function exhibits entanglement, i.e., 
when $\Psi({\bf x}_1,\ldots,{\bf x}_n,t)$ is {\em not}
a local product of the form 
$\psi_1({\bf x}_1,t)\cdots\psi_n({\bf x}_n,t)$, 
then $Q({\bf x}_1,\ldots,{\bf x}_n,t)$ is not 
of the form $\sum_{l=1}^{n}Q_l({\bf x}_l,t)$ (compare with
(\ref{hcl})). 
In the Bohmian interpretation, this means that
$Q$ is the quantum potential
which (in the case of entanglement) describes a nonlocal interaction.  
For attempts to formulate the nonlocal Bohmian interaction 
in a relativistic covariant way, see, e.g., \cite{durpra1,durpra2,
hort,nikolfpl3,nikddw,nikmft}.

\subsection{(Non)locality without hidden variables?}

Concerning the issue of locality, the most difficult question is 
whether QM itself, without hidden variables, is local or not.
The fact is that there is no consensus among experts on that issue.
It is known that quantum effects, such as the Einstein-Podolsky-Rosen-Bell
effect or the Hardy effect, cannot be used to transmit information. 
This is because the choice of the state to which the system 
will collapse is {\em random} (as we have seen, this randomness may
be either fundamental or effective), so one cannot choose to 
transmit the message one wants. In this sense, QM is local.
On the other hand, 
the correlation among different subsystems is nonlocal, in the sense that
one subsystem is correlated with another subsystem, such that this correlation 
cannot be explained in a local manner in terms of preexisting properties
before the measurement.
Thus, there are good reasons for the claim that QM is {\em not} local.

Owing to the nonlocal correlations discussed above, some physicists 
claim that it is a fact that QM is not local. Nevertheless, 
many experts do not agree with this claim, 
so it cannot be regarded as a fact.
Of course, at the conceptual level, it is very difficult 
to conceive how nonlocal correlations can be explained without nonlocality.
Nevertheless, hard-orthodox quantum physicists are trying to do that
(see, e.g., \cite{zeil,rov1,rov2,mermcor}).
In order to save the locality principle, they, 
in one way or another, deny the existence 
of objective reality. Without objective reality, 
there is nothing to be objectively nonlocal.
What remains is the wave function that satisfies a local 
Schr\"odinger equation and does not represent reality, but only 
the information on reality, while reality itself
does not exist in an objective sense. 
Many physicists (including myself) have problems with thinking 
about information on reality without objective reality itself,
but it does not prove that such thinking is incorrect. 

To conclude, the fact is that, so far, there has been no final proof 
with which most experts would agree that QM is either local or 
nonlocal. (For the most recent attempt to establish such a 
consensus see \cite{niknonloc}.) 
There is only agreement that if hidden variables 
(that is, objective physical properties existing even when 
they are not measured) exist, then they must be nonlocal.
Some experts consider this a proof that they do not exist, whereas  
other experts consider this a proof that QM is nonlocal.
They consider these as proofs
because they are reluctant to give up either 
of the principle of locality or of the existence of objective reality. 
Nevertheless, more open-minded (some will say -- too open-minded) 
people admit that neither of these two ``crazy" possibilities 
(nonlocality and absence of objective reality) should be 
{\em a priori} excluded.

\section{There is a well-defined relativistic QM}
\label{RQM}

\subsection{Klein-Gordon equation and the problem of probabilistic 
interpretation}

The free Schr\"odinger equation
\begin{equation}\label{sch}
\frac{-\hbar^2 \nabla^2}{2m} \psi({\bf x},t)=
i\hbar \partial_t \psi({\bf x},t) 
\end{equation}
is not consistent with the theory of relativity. 
In particular, it treats space and time in completely different 
ways, which contradicts the principle of relativistic covariance.
Eq.~(\ref{sch}) 
corresponds only to a nonrelativistic approximation of QM.
What is the corresponding relativistic equation from which 
(\ref{sch}) can be derived as an approximation?
Clearly, the relativistic equation must treat space and time 
on an equal footing. For that purpose, it is convenient 
to choose units in which the velocity of light is $c=1$. 
To further simplify equations, it is also convenient to 
further restrict units so that $\hbar=1$. Introducing
coordinates $x^{\mu}$, $\mu=0,1,2,3$, where $x^0=t$, while 
$x^1,x^2,x^3$ are space coordinates, the simplest
relativistic generalization of (\ref{sch}) is the 
Klein-Gordon equation
\begin{equation}\label{KG}
(\partial^{\mu}\partial_{\mu}+m^2)\psi(x)=0,
\end{equation}
where $x=\{ x^{\mu} \}$, summation over repeated indices is understood,
$\partial^{\mu}\partial_{\mu}=\eta^{\mu\nu}\partial_{\mu}\partial_{\nu}$,
and $\eta^{\mu\nu}$ is the diagonal metric tensor with 
$\eta^{00}=1$, $\eta^{11}=\eta^{22}=\eta^{33}=-1$. 
However, the existence of this relativistic wave equation does {\em not}
imply that relativistic QM exists. This is because there 
are {\em interpretational} problems with this equation.
In nonrelativistic QM, the quantity $\psi^*\psi$ is the probability 
density, having the property 
\begin{equation}\label{prob=1}
\frac{d}{dt}\int d^3x\, \psi^*\psi =0, 
\end{equation}
which can be easily derived from the Schr\"odinger equation (\ref{sch}).
This property is crucial for the consistency of the probabilistic 
interpretation, because the integral
$\int d^3x\, \psi^*\psi$ is the sum of 
all probabilities for the particle to be at all possible places, 
which must be equal to $1$ for each time $t$. If $\psi$ is normalized 
so that this integral is equal to $1$ at $t=0$, then (\ref{prob=1}) 
provides that it is equal to $1$ at each $t$. 
However, when $\psi$ satisfies (\ref{KG}) instead of 
(\ref{sch}), then the consistency requirement (\ref{prob=1}) is not 
fulfilled. Consequently, {\em in relativistic QM based on 
(\ref{KG}), $\psi^*\psi$ cannot be interpreted as the 
probability density}.

In order to solve this problem, one can introduce the 
Klein-Gordon current
\begin{equation}\label{cur}
j_{\mu}=i\psi^* \!\stackrel{\leftrightarrow\;}{\partial_{\mu}}\! \psi ,
\end{equation}
where $a \!\stackrel{\leftrightarrow\;}{\partial_{\mu}}\! b \equiv
a(\partial_{\mu}b) -(\partial_{\mu} a)b$.
Using (\ref{KG}), one can show that this current satisfies the 
local conservation law
\begin{equation}
\partial_{\mu}j^{\mu}=0,
\end{equation}     
which implies that
\begin{equation}\label{prob=1rel}
\frac{d}{dt}\int d^3x\, j^0 =0 .
\end{equation}
Eq. (\ref{prob=1rel}) suggests that, in the relativistc case, 
it is $j^0$ that should be interpreted as the probability density.
More generally, if $\psi_1(x)$ and $\psi_2(x)$ are two solutions
of (\ref{KG}), then the scalar product defined as
\begin{equation}\label{scalprod}
(\psi_1,\psi_2)=i\int d^3x \, \psi_1^*(x)
\!\stackrel{\leftrightarrow\;}{\partial_0}\! \psi_2(x) 
\end{equation} 
does not depend on time. The scalar product (\ref{scalprod})
does not look relativistic covariant, but there is a way to 
write it in a relativistic covariant form. The constant-time
spacelike 
hypersurface with the infinitesimal volume $d^3x$ can be generalized
to an arbitrarily curved spacelike hypersurface $\Sigma$ with
the infinitesimal volume $dS^{\mu}$ oriented in a 
timelike direction normal to $\Sigma$. Eq.~(\ref{scalprod})
then generalizes to  
\begin{equation}\label{scalprod2}
(\psi_1,\psi_2)=i\int_{\Sigma} dS^{\mu} \, \psi_1^*(x)
\!\stackrel{\leftrightarrow\;}{\partial_{\mu}}\! \psi_2(x) , 
\end{equation}  
which, owing to the 4-dimensional Gauss law, does not depend on
$\Sigma$ when $\psi_1(x)$ and $\psi_2(x)$ satisfy (\ref{KG}).
However, there is a problem again. The general solution of 
(\ref{KG}) can be written as 
\begin{equation}\label{e48}
\psi(x)=\psi^+(x) + \psi^-(x),
\end{equation}
where
\begin{eqnarray}\label{psi+-}
& \psi^+(x)=
\displaystyle\sum_{{\bf k}} 
c_{{\bf k}} e^{-i(\omega_{{\bf k}}t-{\bf k}{\bf x}) } , &
\\ \nonumber
& \psi^-(x)=
\displaystyle\sum_{{\bf k}} 
d_{{\bf k}} e^{i(\omega_{{\bf k}}t-{\bf k}{\bf x}) } . &
\end{eqnarray} 
Here $c_{{\bf k}}$ and $d_{{\bf k}}$ are arbitrary complex coefficients,
and
\begin{equation}
\omega_{{\bf k}}\equiv\sqrt{{\bf k}^2+m^2} 
\end{equation}
is the frequency.
For obvious reasons, $\psi^+$ is called a {\em positive-frequency} solution, 
while $\psi^-$ is called a {\em negative-frequency} solution.
(The positive- and negative-{\em frequency} solutions are often referred to as 
positive- and negative-{\em energy} solutions, respectively. 
However, such a terminology is misleading because in field theory, 
which will be discussed in the next section, energy cannot be negative, 
so it is better to speak of positive and negative frequency.)
The nonrelativistic Schr\"odinger equation contains only 
the positive-frequency solutions, which can be traced back 
to the fact that the Schr\"odinger equation contains a first time 
derivative, instead of a second time derivative that appears 
in the Klein-Gordon equation (\ref{KG}).
For a positive-frequency solution the quantity $\int d^3x\, j^0$ is
positive, whereas for a negative-frequency solution this quantity 
is negative. Since the sum of all probabilities must be positive, 
the negative-frequency solutions represent a problem for the 
probabilistic interpretation. One may propose that only 
positive-frequency solutions are physical, but even this does not 
solve the problem. Although the integral $\int d^3x\, j^0$ is 
strictly positive in that case, the local density $j^0(x)$ may still be 
negative at some regions of spacetime, provided that the 
superposition $\psi^+$ in (\ref{psi+-}) contains 
terms with two or more different positive frequencies.
Thus, even with strictly positive-frequency solutions, the 
quantity $j^0$ cannot be interpreted as a probability density.  
 
\subsection{Some attempts to solve the problem}

Physicists sometimes claim that there are no interpretational 
problems with the Klein-Gordon equation because the coefficients 
$c_{{\bf k}}$ and $d_{{\bf k}}$ in (\ref{psi+-}) (which are 
the Fourier transforms of $\psi^+$ and $\psi^-$, respectively) 
are time independent, so the quantities $c^*_{{\bf k}}c_{{\bf k}}$
and $d^*_{{\bf k}}d_{{\bf k}}$ can be consistently interpreted as 
probability densities in the momentum space. 
(More precisely, if $c_{{\bf k}}$ and $d_{{\bf k}}$ are independent,
then these two probability densities refer to particles and 
antiparticles, respectively.) Indeed, in practical 
applications of relativistic QM, one is often interested only in 
scattering processes, in which the probabilities of different 
momenta contain all the information that can be compared 
with actual experiments. From a practical point of view, this is 
usually enough. Nevertheless, in principle, it is possible  
to envisage an experiment in which one measures the probabilities
in the position (i.e., configuration) space, rather than that 
in the momentum space. A complete theory should have predictions on 
all quantities that can be measured in principle.
Besides, if the standard interpretation of the nonrelativistic wave function 
in terms of the probability density in the position space is 
correct (which, indeed, is experimentally confirmed), then this 
interpretation must be derivable from a more accurate theory -- 
relativistic QM. Thus, the existence of the probabilistic interpretation 
in the momentum space does not really solve the problem.

It is often claimed that the problem of relativistic 
probabilistic interpretation in the position space is solved 
by the Dirac equation. As we have seen, the problems with the 
Klein-Gordon equation can be traced back to the fact that it 
contains a second time derivative, instead of a first one.
The relativistic-covariant wave equation that contains only 
first derivatives with respect to time and space is the
Dirac equation
\begin{equation}\label{dirac}
(i\gamma^{\mu}\partial_{\mu}-m)\psi(x)=0 .
\end{equation}
Here $\gamma^{\mu}$ are the $4\times 4$ Dirac matrices 
that satisfy the anticommutation relations
\begin{equation}\label{anticom}
\{ \gamma^{\mu}, \gamma^{\nu} \}=2\eta^{\mu\nu} ,
\end{equation}            
where $\{A,B\}\equiv AB+BA$.
The Dirac matrices are related to the Pauli matrices $\sigma_i$ 
discussed in Sec.~\ref{NOREAL}, which satisfy 
$\{ \sigma_i, \sigma_j \}=2\delta_{ij}$. (For more details, 
see, e.g., \cite{BD1}.) It turns out that $\psi$ in (\ref{dirac}) 
is a 4-component wave function called spinor that describes particles 
with spin $\frac{1}{2}$. The conserved current associated with 
(\ref{dirac}) is    
\begin{equation}\label{curdir}
j^{\mu}=\bar{\psi}\gamma^{\mu}\psi ,
\end{equation}
where $\bar{\psi}\equiv \psi^{\dagger}\gamma^0$. 
In particular, (\ref{anticom}) implies $\gamma^0\gamma^0=1$, 
so (\ref{curdir}) implies
\begin{equation}
j^0=\psi^{\dagger}\psi ,
\end{equation}
which cannot be negative. Thus, the Dirac equation does not 
have problems with the probabilistic interpretation. 
However, this still does not mean that the problems of 
relativistic QM are solved. This is because the Dirac equation 
describes only particles with spin $\frac{1}{2}$. 
Particles with different spins also exist in nature.
In particular, the Klein-Gordon equation describes particles with  
spin $0$, while the wave equation for spin $1$ particles are 
essentially the Maxwell equations, which are second-order 
differential equations for the electromagnetic potential 
$A^{\mu}$ and lead to the same interpretational problems 
as the Klein-Gordon equation.   

There are various proposals for a more direct solution to the problem of 
probabilistic interpretation of the Klein-Gordon equation
(see, e.g., \cite{newt,ghose,gavr,nikolfpl3,nikolfol}). 
However, all these proposed solutions 
have certain disadvantages and none of these proposals 
is widely accepted as the correct solution.
Therefore, without this problem being 
definitely solved, it cannot be said that there exists a 
well-defined relativistic QM.   

\section{Quantum field theory solves the problems of relativistic QM}
\label{QFT}

It is often claimed that the interpretational problems 
with relativistic QM discussed in the preceding section 
are solved by a more advanced theory -- {\em quantum field theory} (QFT).
To see how QFT solves these problems and whether this solution 
is really satisfactory, let me briefly review 
what QFT is and why it was introduced.

\subsection{Second quantization of particles}

A theoretical concept closely related to QFT is the 
{\em method of second quantization}. It was introduced 
to formulate in a more elegant way the fact that 
many-particle wave functions should be either completely symmetric
or completely antisymmetric under exchange of any two particles, 
which comprises the principle that identical particles 
cannot be distinguished. Let
\begin{equation}\label{wf1}
\psi({\bf x},t)=\sum_k a_k f_k({\bf x},t)
\end{equation}
be the wave function expanded in terms of some 
complete orthonormal set of solutions $f_k({\bf x},t)$. 
(For free particles, $f_k({\bf x},t)$ are usually taken to 
be the plane waves $f_k({\bf x},t) \propto e^{-i(\omega t - {\bf kx})}$.)
Unlike the particle position ${\bf x}$, the wave function $\psi$
does not correspond to an operator. Instead, it is just 
an ordinary number that determines the probability density
$\psi^*\psi$. This is so in the ordinary ``first" quantization 
of particles. The method of second quantization promotes 
the wave function $\psi$ to an operator $\hat{\psi}$. 
(To avoid confusion, from now on, 
the operators are always denoted by a hat above it.)
Thus, instead of (\ref{wf1}), we have the operator
\begin{equation}\label{wf2}
\hat{\psi}({\bf x},t)=\sum_k \hat{a}_k f_k({\bf x},t) ,
\end{equation}   
where the coefficients $a_k$ are also promoted to the 
operators $\hat{a}_k$. Similarly, instead of the complex conjugated 
wave function $\psi^*$, we have the hermitian conjugated 
operator
\begin{equation}\label{wf2*}              
\hat{\psi}^{\dagger}({\bf x},t)=\sum_k \hat{a}^{\dagger}_k f_k^*({\bf x},t) .
\end{equation}
The orthonormal solutions $f_k({\bf x},t)$ are still ordinary functions 
as before, so that the operator $\hat{\psi}$ satisfies the same equation of 
motion (e.g., the Schr\"odinger equation in the nonrelativistic case)
as $\psi$. In the case of bosons, 
the operators $\hat{a}_k,\hat{a}_k^{\dagger}$ are postulated to satisfy the 
commutation relations
\begin{eqnarray}\label{combos}
& [\hat{a}_k,\hat{a}_{k'}^{\dagger}]=\delta_{kk'} , & \nonumber \\
& [\hat{a}_k,\hat{a}_{k'}]=
[\hat{a}_k^{\dagger},\hat{a}_{k'}^{\dagger}]=0 .
\end{eqnarray}
These commutation relations are postulated because,
as is well-known from the case of first-quantized harmonic oscillator
(discussed also in more detail in the next section),
such commutation relations lead to a representation in which
$\hat{a}_k^{\dagger}$ and $\hat{a}_k$ are raising and lowering operators,
respectively. Thus, an $n$-particle state with the 
wave function $f(k_1,\ldots,k_n)$ in the 
$k$-space can be abstractly represented as
\begin{equation}\label{nf}
|n_f\rangle =\sum_{k_1,\ldots,k_n} f(k_1,\ldots,k_n) \, 
\hat{a}^{\dagger}_{k_1} \cdots \hat{a}^{\dagger}_{k_n} |0\rangle ,
\end{equation}
where $|0\rangle$ is the ground state
of second quantization, i.e., the vacuum state 
containing no particles. 
Introducing the operator
\begin{equation}\label{N}
\hat{N}=\sum_{k} \hat{a}^{\dagger}_k \hat{a}_k ,
\end{equation}
and using (\ref{combos}), one can show that
\begin{equation}\label{N2}
\hat{N}|n_f\rangle = n |n_f\rangle .
\end{equation}
Since $n$ is the number of particles in the state (\ref{nf}),
Eq.~(\ref{N2}) shows that $\hat{N}$ is the operator of the 
number of particles.
The $n$-particle wave function 
in the configuration space can then be written as 
\begin{equation}\label{wfn}
\psi({\bf x}_1,\ldots,{\bf x}_n, t)=
\langle 0|\hat{\psi}({\bf x}_1,t)\cdots\hat{\psi}({\bf x}_n,t)|n_f\rangle .
\end{equation}
From (\ref{combos}) and (\ref{wf2}) we see that 
$\hat{\psi}({\bf x},t)\hat{\psi}({\bf x}',t)=
\hat{\psi}({\bf x}',t)\hat{\psi}({\bf x},t)$, which implies 
that the ordering of the $\hat{\psi}$-operators in (\ref{wfn}) is 
irrelevant. This means that (\ref{wfn}) automatically 
represents a bosonic wave function completely symmetric under 
any two exchanges of the arguments ${\bf x}_a$, $a=1,\ldots,n$.
For the fermionic case, one replaces the commutation relations
(\ref{combos}) with similar anticommutation relations  
\begin{eqnarray}\label{comfer}
& \{\hat{a}_k,\hat{a}_{k'}^{\dagger}\}=\delta_{kk'} , & \nonumber \\
& \{\hat{a}_k,\hat{a}_{k'}\}=
\{\hat{a}_k^{\dagger},\hat{a}_{k'}^{\dagger}\}=0 ,
\end{eqnarray}
which, in a similar way, leads to completely antisymmetric
wave functions.

\subsection{Quantum fields}

The method of second quantization outlined above 
is nothing but a convenient mathematical trick.
It does not bring any new physical information. However, 
the mathematical formalism used in this trick 
can be {\em reinterpreted} in the following way:
The fundamental quantum object is neither the particle 
with the position-operator $\hat{{\bf x}}$ nor the 
wave function $\psi$, but a new {\em hermitian operator}
\begin{equation}\label{field}
\hat{\phi}({\bf x},t) =
\hat{\psi}({\bf x},t) + \hat{\psi}^{\dagger}({\bf x},t) .
\end{equation}
This hermitian operator is called {\em field} and 
the resulting theory is called {\em quantum field theory} (QFT).
It is a quantum-operator version of a {\em classical} field 
$\phi({\bf x},t)$. 
(A prototype of classical fields 
is the electromagnetic field satisfying Maxwell equations.
Here, for pedagogical purposes, we do not study the 
electromagnetic field, but only the simplest scalar field $\phi$.) 
Using (\ref{wf2}), (\ref{wf2*}), and (\ref{combos}), one obtains
\begin{eqnarray}
[\hat{\phi}({\bf x},t),\hat{\phi}({\bf x}',t)] & = &
\sum_k f_k({\bf x},t) f_k^*({\bf x}',t) \nonumber \\
& & -\sum_k f_k^*({\bf x},t) f_k({\bf x}',t) .
\end{eqnarray}
Thus, by
using the completeness relations
\begin{equation}
\sum_k f_k({\bf x},t) f_k^*({\bf x}',t)=
\sum_k f_k^*({\bf x},t) f_k({\bf x}',t)=\delta^3({\bf x}-{\bf x}') ,
\end{equation}
one finally obtains
\begin{equation}\label{comf}
[\hat{\phi}({\bf x},t),\hat{\phi}({\bf x}',t)]=0.
\end{equation}
Thus, from (\ref{wf2}), (\ref{wf2*}), (\ref{combos}), and (\ref{field}) one 
finds that (\ref{wfn}) can also be written as
\begin{equation}\label{wfn2}
\psi({\bf x}_1,\ldots,{\bf x}_n, t)=
\langle 0|\hat{\phi}({\bf x}_1,t)\cdots\hat{\phi}({\bf x}_n,t)|n_f\rangle ,
\end{equation}
which, owing to (\ref{comf}), provides the complete symmetry
of $\psi$. 

The field equations of motion are derived from their own
actions. For example, the Klein-Gordon equation (\ref{KG})
for $\phi(x)$ (instead of $\psi(x)$)
can be obtained from the classical action
\begin{equation} 
A=\int d^4x {\cal L} ,
\end{equation}
where
\begin{equation}\label{lagr}
{\cal L}(\phi,\partial_{\alpha}\phi)=
\frac{1}{2} [(\partial^{\mu}\phi)(\partial_{\mu}\phi)
-m^2\phi^2 ]    
\end{equation}
is the Lagrangian density. The canonical momentum associated
with this action is a fieldlike quantity
\begin{equation}
\pi(x)=\frac{\partial{\cal L}}{\partial (\partial_0\phi(x))}
=\partial^0\phi(x) .
\end{equation}
The associated Hamiltonian density is
\begin{equation}\label{hamden}
{\cal H}=\pi \partial_0\phi - {\cal L}=
\frac{1}{2} [\pi^2 + (\nabla\phi)^2 +m^2\phi^2] .
\end{equation}
This shows that the field energy 
\begin{equation}\label{hamfi}
H[\pi,\phi]=\int d^3x\, {\cal H}(\pi({\bf x}),\phi({\bf x}),
\nabla\phi({\bf x}) ) 
\end{equation}
(where the time-dependence is suppressed)
cannot be negative. 
This is why, in relativistic QM, it is better to speak
of negative frequencies than of negative energies.
(In (\ref{hamfi}), the notation $H[\pi,\phi]$ denotes 
that $H$ is {\em not} a function of 
$\pi({\bf x})$ and $\phi({\bf x})$ at some particular values of 
${\bf x}$, but a {\em functional}, i.e., an object that 
depends on the {\em whole} functions $\pi$ and $\phi$ at
{\em all} values of ${\bf x}$.) 
By analogy with the particle commutation relations
$[\hat{x}_l,\hat{p}_m]=i\delta_{lm}$, 
$[\hat{x}_i,\hat{x}_j]=[\hat{p}_i,\hat{p}_j]=0$, 
the fundamental field-operator commutation relations are
postulated to be
\begin{eqnarray}\label{comfund}
& [\hat{\phi}({\bf x}),\hat{\pi}({\bf x}')]=i\delta^3({\bf x}-{\bf x}') , &
\nonumber \\
& [\hat{\phi}({\bf x}),\hat{\phi}({\bf x}')]=
[\hat{\pi}({\bf x}),\hat{\pi}({\bf x}')]=0 .
\end{eqnarray} 
Here it is understood that all fields are evaluated at the same time $t$, 
so the $t$ dependence is not written explicitly.
Thus, now (\ref{comf}) is one of the fundamental (not derived) 
commutation relations. Since $\hat{\phi}(x)$ 
is an operator in the Heisenberg picture that satisfies the 
Klein-Gordon equation, the expansion (\ref{field}) with 
(\ref{wf2}) and (\ref{wf2*}) can be used. One of the most important 
things gained from quantization of fields is the fact that 
now the commutation relations (\ref{combos}) do {\em not} need
to be postulated. Instead, they can be derived from 
the fundamental field-operator commutation relations 
(\ref{comfund}). (The fermionic field-operators satisfy similar 
fundamental relations with commutators
replaced by anticommutators, from which 
(\ref{comfer}) can be derived.)
The existence of the Hamiltonian (\ref{hamfi}) allows us 
to introduce the functional Schr\"odinger equation
\begin{equation}\label{funcsch}
H[\hat{\pi},\phi] \Psi[\phi;t)=i\partial_t \Psi[\phi;t) ,
\end{equation}
where
$\Psi[\phi;t)$ is a functional of $\phi({\bf x})$ and a function of 
$t$, while 
\begin{equation}
\hat{\pi}({\bf x}) = -i\frac{\delta}{\delta\phi({\bf x})}
\end{equation}
is the field analog of the particle-momentum operator
$\hat{p}_j=-i\partial/\partial x_j$. 
(For a more careful definition of the functional 
derivative $\delta/\delta\phi({\bf x})$ see, e.g., \cite{ryder}.)
Unlike the Klein-Gordon equation, the functional 
Schr\"odinger equation (\ref{funcsch}) is a first-order differential 
equation in the time derivative. Consequently, the quantity
\begin{equation}\label{rhofi}
\rho[\phi;t)=\Psi^*[\phi;t)\Psi[\phi;t) 
\end{equation}
can be consistently interpreted as a conserved probability density.
It represents the probability that the field has the 
configuration $\phi({\bf x})$ at the time $t$.

\subsection{Does QFT solve the problems of relativistic QM?}

After this brief overview of QFT, we are finally ready 
to cope with the validity of the title of this section. 
How QFT helps in solving the interpretational problems 
of relativistic QM? According to QFT, the fundamental 
objects in nature are not particles, but fields.
Consequently, the fundamental wave function(al) that
needs to have a well-defined probabilistic interpretation 
is not $\psi({\bf x},t)$, but $\Psi[\phi;t)$. 
Thus, the fact that, in the case of Klein-Gordon equation,
$\psi({\bf x},t)$ cannot be interpreted 
probabilistically, is no longer a problem from this more 
fundamental point of view. However, does it really solve 
the problem? If QFT is really a more fundamental 
theory than the first-quantized quantum theory of 
particles, then it should be able to reproduce {\em all}
good results of this less fundamental theory. 
In particular, from the fundamental 
axioms of QFT (such as the axiom that (\ref{rhofi}) represents 
the probability in the space of fields), one should be able to deduce 
that, at least in the nonrelativistic limit, $\psi^*\psi$ represents 
the probability in the space of particle positions. 
However, one {\em cannot} deduce it solely from the axioms of 
QFT. One possibility is to 
completely ignore, or even deny \cite{zeh}, the validity of the 
probabilistic interpretation of 
$\psi$, which indeed is in the spirit of QFT viewed as a fundamental 
theory, but then the problem is to reconcile it with the 
fact that such a probabilistic interpretation of $\psi$
is in agreement with experiments. Another possibility 
is to supplement the axioms of QFT with an additional 
axiom that says that $\psi$ in the nonrelativistic limit 
determines the probabilities of particle positions, 
but then such a set of axioms is not coherent, as it does not 
specify the meaning of $\psi$ in the relativistic case.
Thus, instead of saying that QFT solves the problems of 
relativistic QM, it is more honest to say that it merely sweeps 
them under the carpet.            
 
\section{Quantum field theory is a theory of particles}
\label{QFTP}

What is the world made of? A common answer is that it is made 
of {\em elementary particles}, such as electrons, photons, quarks, 
gluons, etc. On the other hand, all modern theoretical 
research in elementary-particle physics is based on 
quantum {\em field} theory (QFT) \cite{BD2,ryder,cheng}. 
So, is the world made of 
particles or fields? A frequent answer given by 
elementary-particle physicists is that QFT is actually
a theory of particles, or more precisely, that particles are actually more 
fundamental physical objects, while QFT is more like a mathematical tool 
that describes -- the particles. Indeed, the fact that the motivation 
for introducing QFT partially emerged from the method of second
quantization (see Sec.~\ref{QFT}) supports this interpretation 
according to which QFT is nothing but a theory of particles. 
But is that really so?
Is it really a fundamental property of QFT that it describes particles?
Let us see! 

\subsection{A first-quantized analog of particles in QFT}

From the conceptual point of view, fields and particles are very 
different objects. This is particularly clear for classical 
fields and particles, where all concepts are clear. So, if there exists 
a relation between {\em quantum} fields and particles that does not 
have an analog in the classical theory of fields and particles, 
then such a relation must be highly nontrivial. Indeed, this 
nontrivial relation is related to the nontrivial commutation relations
(\ref{combos}) (or (\ref{comfer}) for fermionic fields). 
The classical fields commute, which implies that 
the classical coefficients $a_k$, $a_k^*$ do not satisfy (\ref{combos}).
Without these commutation relations, we could {\em not}
introduce $n$-particle states (\ref{nf}). 
However, are the commutation relations (\ref{combos}) sufficient 
for having a well-defined notion of particles? To answer this 
question, it is instructive to study the analogy with 
the first-quantized theory of particles. 

Consider a quantum particle 
moving in one dimension, having a Hamiltonian
\begin{equation}\label{ham1}
\hat{H}=\frac{\hat{p}}{2m}+V(\hat{x}) .
\end{equation}
We introduce the operators
\begin{eqnarray}\label{a1p}
& \hat{a}=\frac{1}{\sqrt{2}} \left( \sqrt{m\omega}\hat{x} 
+i \displaystyle\frac{\hat{p}}{\sqrt{m\omega}} \right) , & \nonumber \\
& \hat{a}^{\dagger}=
\frac{1}{\sqrt{2}} \left( \sqrt{m\omega}\hat{x}   
-i \displaystyle\frac{\hat{p}}{\sqrt{m\omega}} \right) ,
\end{eqnarray}
where $\omega$ is some constant of the dimension of energy
(or frequency, which, since $\hbar=1$, has the same dimension as energy).
Using the commutation relation $[\hat{x},\hat{p}]=i$, we obtain
\begin{equation}
[\hat{a},\hat{a}^{\dagger}]=1 .
\end{equation}
This, together with the trivial commutation relations
$[\hat{a},\hat{a}]=[\hat{a}^{\dagger},\hat{a}^{\dagger}]=0$, 
shows that $\hat{a}^{\dagger}$ and $\hat{a}$ are the 
raising and lowering operator, respectively.
As we speak of {\em one} particle, 
the number operator $\hat{N}=\hat{a}^{\dagger}\hat{a}$ 
now cannot be called the number of particles. Instead, 
we use a more general terminology (applicable to (\ref{N}) 
as well) according to which $\hat{N}$ is the number of 
``quanta". But quanta of what? It is easy to show that 
(\ref{ham1}) can be written as
\begin{equation}\label{ham2}
\hat{H}=\omega\left( \hat{N}+\frac{1}{2} \right)
+\left[ V(\hat{x})-\frac{m\omega^2\hat{x}^2}{2} \right] .
\end{equation}  
In the special case in which 
$V(\hat{x})=m\omega^2\hat{x}^2/2$, which corresponds to the 
harmonic oscillator, the square bracket in (\ref{ham2})
vanishes, so the Hamiltonian can be expressed 
in terms of the $\hat{N}$-operator only. In this case, 
the (properly normalized) state
\begin{equation}\label{staten}
|n\rangle = \frac{(\hat{a}^{\dagger})^n}{\sqrt{n!}} |0\rangle
\end{equation}
has the energy $\omega (n+1/2)$, so the energy can be viewed
as a sum of the ground-state energy $\omega/2$ and the
energy of {\em $n$ quanta with energy $\omega$}. 
This is why the number operator $\hat{N}$ plays an important physical role.  
However, the main point that can be inferred from (\ref{ham2}) 
is the fact that, for a {\em general} potential $V(\hat{x})$, 
{\em the number operator
$\hat{N}$ does not play any particular physical role}.
Although the spectrum of quantum states can often be labeled 
by a discrete label $n'=0,1,2,\ldots$, this label, in general, 
has nothing to do with the operator $\hat{N}$ (i.e., the eigenstates
(\ref{staten}) of $\hat{N}$ are not the eigenstates of 
the Hamiltonian (\ref{ham2})). Moreover, 
in general, the spectrum of energies $E(n')$ may have a more 
complicated dependence on $n'$, so that, unlike the 
harmonic oscillator, the spectrum of energies is {\em not equidistant}.
Thus, in general, the state of a system 
{\em cannot} be naturally specified by a number of ``quanta" $n$.  

If one insists on representing the system in terms of the states 
(\ref{staten}), then one can treat the square bracket in (\ref{ham2}) 
as a perturbation $V_{\rm I}(\hat{x})$. (Here ``I" stands for
``interaction".) 
From (\ref{a1p}) one finds 
\begin{equation}\label{e81}
\hat{x}=\frac{\hat{a}+\hat{a}^{\dagger}}{\sqrt{2m\omega}} ,
\end{equation}
so $V_{\rm I}(\hat{x})=V_{\rm I}(\hat{a},\hat{a}^{\dagger})$.
Consequently,
various terms in the perturbation expansion can be represented 
in terms of creation and destruction of quanta, owing to
the occurrence of $\hat{a}^{\dagger}$ and $\hat{a}$, respectively.
However, treating the square bracket in (\ref{ham2}) as a 
perturbation is completely arbitrary. Such a treatment 
is nothing but a mathematical convenience and does not make 
the states (\ref{staten}) more physical. This is particularly 
clear for the cases in which the original system with the 
Hamiltonian (\ref{ham1}) can be solved analytically, without 
a perturbation expansion. The creation and destruction of 
quanta appearing in the perturbation expansion does not
correspond to actual physical processess. 
These ``processess" of creation and destruction are nothing 
but a verbalization of certain mathematical terms appearing {\em only} 
in one particular method of calculation -- the perturbation expansion 
with the square bracket in (\ref{ham2}) treated as 
the perturbation. Last but not least, even if, despite the unnaturalness, 
one decides to express everything in terms of the operators
(\ref{a1p}) and the states (\ref{staten}), there still may remain an 
ambiguity in choosing the constant $\omega$.   
All this demonstrates that, in general, {\em QM
is not a theory of ``quanta" attributed to the operator $\hat{N}$.}

\subsection{Particles in perturbative QFT}

The analogy between the notion of ``quanta" in the first-quantized 
theory of particles and the notion of ``particles" in QFT is 
complete. For example, the QFT analog of (\ref{e81}) is the 
field operator in the Schr\"odinger picture
\begin{equation}
\hat{\phi}({\bf x})=\sum_k \hat{a}_k f_k({\bf x}) + 
\hat{a}_k^{\dagger} f_k^*({\bf x}) ,
\end{equation}
which corresponds to (\ref{field}) with (\ref{wf2}) and (\ref{wf2*}), 
at fixed $t$. If $f_k({\bf x},t)$ are the plane waves proportional 
to $e^{-i(\omega_{\bf k}t-{\bf k}{\bf x})}$, then 
the quantum Hamiltonian obtained from the 
Lagrangian density (\ref{lagr}) turns out to be
\begin{equation}\label{hamfield}
\hat{H}=\sum_{\bf k} \omega_{\bf k}\left(\hat{N}_{\bf k} +\frac{1}{2}
\right) ,
\end{equation}
with $\hat{N}_{\bf k}\equiv \hat{a}_{\bf k}^{\dagger} \hat{a}_{\bf k}$, 
which is an analog of the first term in (\ref{ham2}). 
This analogy is related to the fact that (\ref{lagr}) represents 
a relativistic-field generalization of the harmonic oscillator.
(The harmonic-oscillator Lagrangian is quadratic in 
${\bf x}$ and its derivative, while (\ref{lagr}) is quadratic 
in $\phi$ and its derivatives).
The Hamiltonian (\ref{hamfield}) has a clear physical interpretation;
ignoring the term $1/2$ (which corresponds to an irrelevant
ground-state energy $\sum_{\bf k} \omega_{\bf k}/2$),
%a single particle may have an arbitrary energy $\omega_{\bf k}$, 
for each $\omega_{\bf k}$ there can be only an {\em integer}
number $n_{\bf k}$ of quanta with energy $\omega_{\bf k}$, so that 
their total energy sums up to $n_{\bf k}\omega_{\bf k}$.
(Concerning the irrelevance of the ground-state energy above,
it should be noted that it is often claimed that this energy is relevant 
for the experimentally confirmed Casimir effect, 
but that the fact is that this effect 
can be derived even without referring to the ground-state energy
\cite{jaffe}.) 
These quanta are naturally interpreted as ``particles" with  
energy $\omega_{\bf k}$.  
However, the Lagrangian (\ref{lagr}) is only a special case.
In general, a Lagrangian describing the field $\phi$ may have a form
\begin{equation}\label{lagr2}
{\cal L}=
\frac{1}{2} (\partial^{\mu}\phi)(\partial_{\mu}\phi) -V(\phi) ,
\end{equation} 
where $V(\phi)$ is an arbitrary potential. Thus, in general, 
the Hamiltonian contains an additional term analogous to that in 
(\ref{ham2}), which destroys the ``particle"-interpretation of the 
spectrum.

Whereas the formal mathematical analogy between first quantization and 
QFT (which implies the irrelevance of the number operator $\hat{N}$) 
is clear, 
there is one crucial {\em physical} difference: Whereas in first quantization
there is really no reason to attribute a special meaning 
to the operator $\hat{N}$, there is an {\em experimental} evidence 
that this is not so for QFT. The existence of 
particles is an {\em experimental fact}! Thus, if one wants to describe 
the experimentally observed objects, one must either reject QFT
(which, indeed, is what many elementary-particle physicists were doing 
in the early days of elementary-particle physics and some of them are 
doing it even today \cite{schub}),
or try to artificially adapt QFT such that it 
remains a theory of particles even with general interactions 
(such as those in (\ref{lagr2})). From a pragmatic and phenomenological 
point of view, the latter strategy turns out to be surprisingly 
successful! For example, in the case of (\ref{lagr2}),
one artificially defines the interaction part of the Lagrangian as
\begin{equation}
V_{\rm I}(\phi)=V(\phi)-\frac{1}{2}m^2\phi^2 ,
\end{equation}
and treats it as a perturbation of the ``free" Lagrangian (\ref{lagr}).
For that purpose, it is convenient to introduce a mathematical trick
called {\em interaction picture}, which is
a picture that interpolates between 
the Heisenberg picture (where the time dependence is attributed 
to fields $\phi$) and the Schr\"odinger picture 
(where the time dependence is
attributed to states $|\Psi\rangle$). In the interaction picture,
the field satisfies the {\em free} Klein-Gordon equation of motion,
while the time evolution of the state is governed only by the interaction 
part of the Hamiltonian. This trick allows one to use
the free expansion (\ref{field}) with (\ref{wf2}) and (\ref{wf2*}), 
despite the fact that the ``true" quantum operator $\hat{\phi}({\bf x},t)$
in the Heisenberg picture cannot be expanded in that way.
In fact, {\em all} operators in the interaction picture 
satisfy the free equations of motion, so the particle-number operator
can also be introduced in the same way as for free fields.   
Analogously to the case of first quantization discussed after
Eq.~(\ref{e81}), certain mathematical terms in the perturbation expansion 
can be pictorially represented by the so-called {\em Feynman diagrams}.
(For technical details, I refer the reader to \cite{BD2,cheng}.)
In some cases, the final measurable results obtained in that way turn out 
to be in excellent agreement with experiments. 

\subsection{Virtual particles?}

The calculational tool represented by Feynman diagrams  
suggests an often abused picture according to which
``real particles interact by exchanging virtual particles".
Many physicists, especially nonexperts,
take this picture literally, as something that 
really and objectively happens in nature. In fact, I have 
{\em never} seen 
a popular text on particle physics in which this picture was 
{\em not} presented as something that really happens.
Therefore, this picture of quantum interactions as processes  
in which virtual particles exchange is one of the 
most abused myths, not only in quantum physics, but in 
physics in general. Indeed, there is a consensus among experts 
for foundations of QFT that such a picture should
not be taken literally. The fundamental principles 
of quantum theory do not even contain a notion of a
``virtual" state. The notion of a 
``virtual particle" originates {\em only} from a 
specific mathematical method of calculation, called perturbative 
expansion. In fact, perturbative expansion 
represented by Feynman diagrams can be introduced even in 
{\em classical} physics \cite{thorn,penco}, but nobody 
attempts to verbalize these classical Feynman diagrams 
in terms of classical ``virtual" processes.  
So why such a verbalization is tolerated in quantum physics?
The main reason is the fact that the standard interpretation 
of quantum theory does not offer a clear ``canonical" ontological picture 
of the actual processes in nature, but only provides 
the probabilities for the final results of measurement outcomes.
In the absence of such a ``canonical" picture, 
physicists take the liberty to introduce 
various auxiliary intuitive pictures that sometimes help them 
think about otherwise abstract quantum formalism. Such auxiliary
pictures, by themselves, are not a sin. However, a potential 
problem occurs when one forgets why such a picture has been introduced 
in the first place and starts to think on it too literally.   

\subsection{Nonperturbative QFT}

In some cases, the picture of particles suggested by the 
``free" part of the Lagrangian does not really correspond 
to particles observed in nature. The best known 
example is {\em quantum chromodynamics} (QCD), a QFT theory 
describing strong interactions between quarks and gluons.
In nature we do not observe quarks, but rather more complicated
particles called {\em hadrons} (such as protons, neutrons, 
and pions). In an oversimplified but often abused picture, 
hadrons are built of 2 or 3 quarks glued together
by gluons. However, since free quarks are never observed in nature,
the perturbative expansion, so successful for some other 
QFT theories, is not very successful in the case of QCD.
Physicists are forced to develop other approximative 
methods to deal with it. The most successful such method 
is the so-called {\em lattice} QCD (for an introductory 
textbook see \cite{creutz} and for pedagogic reviews
see \cite{davies,sharpe}). In this method, the spacetime 
continuum is approximated by a finite lattice
of spacetime points, allowing the application 
of brutal-force numerical methods of computation.
This method allows to compute the expectation values of 
products of fields in the ground state, by starting from first 
principles. However, to extract the information about 
particles from these purely field-theoretic quantities, one must 
{\em assume} a relation between these expectation values
and the particle quantities. This relation is not derived 
from lattice QCD itself, but rather from the known 
relation between fields and particles in perturbative QFT.  
Consequently, although this method reproduces the experimental
hadron data more-or-less successfully, 
the concept of particle in this method is 
not more clear than that in the perturbative approach.
Thus, the notion of real particles
is not derived from first principles
%are introduced in a theoretically {\em ad hoc} (but experimentally 
%justified!) manner, 
and nothing in the formalism 
suggests a picture of the ``exchange of virtual particles".         

\subsection{Particles and the choice of time}

As we have seen,
although the notion of particles in interacting QFT theories 
cannot be derived from first principles (or at least 
we do not know yet how to do that), there are heuristic    
mathematical procedures that introduce the notion of particles 
that agrees with experiments. However, there are circumstances
in QFT in which the theoretical notion of particles is even more ambiguous, 
while present experiments are not yet able to resolve these ambiguities.
Consider again the {\em free} field expanded as in 
(\ref{field}) with (\ref{wf2}) and (\ref{wf2*}).  
The notion of particles rests on a clear separation
between the creation operators $a_k^{\dagger}$ and the destruction 
operators $a_k$. The definition of these operators is 
closely related to the choice of the complete orthonormal 
basis $\{ f_k({\bf x},t) \}$ of solutions to the classical 
Klein-Gordon equation. However, there are 
infinitely many different choices of this basis. 
The plane-wave basis 
\begin{equation}\label{pwb}
f_{\bf k}({\bf x},t) \propto e^{-i(\omega_{\bf k}t-{\bf k}{\bf x})}
\equiv e^{-ik\cdot x} 
\end{equation}
is only a particular convenient choice. Different choices may lead 
to different creation and destruction operators $a_k^{\dagger}$ and 
$a_k$, and thus to different notions of particles. 
How to know which choice is the right one? Eq.~(\ref{pwb}) 
suggests a physical criterion according to which 
the modes $f_k({\bf x},t)$ should be chosen such that 
they have a positive frequency. However, the notion of 
frequency assumes the notion of time. On the other hand, 
according to the theory of relativity, there is not 
a unique choice of the time coordinate. Therefore, 
the problem of the right definition of particles reduces 
to the problem of the right definition of time.
Fortunately, the last exponential function in (\ref{pwb}) shows that 
the standard plane waves $f_{\bf k}(x)$ are Lorentz invariant, 
so that different time coordinates related by a Lorentz 
transformation lead to the same definition of particles.
However, Lorentz transformations relate only proper coordinates 
attributed to inertial observers in flat spacetime.
The {\em general} theory of relativity allows much more 
general coordinate transformations, such as those that 
relate an inertial observer with an accelerating one.
(For readers who are not familiar with general 
theory of relativity there are many excellent introductory
textbooks, but my favored one that I highly recommend 
to the beginners is \cite{carrol}. For an explicit 
construction of the coordinate transformations between 
an inertial observer and an arbitrarily moving one 
in flat spacetime, see \cite{nels,nikpra}, and for instructive applications, 
see \cite{nikpra,nikajp,niktwin}.
Nevertheless, to make this paper readable by those who are not 
familiar with general relativity, in the rest of Sec.~\ref{QFTP}, 
as well as in Sec.~\ref{BH}, I omit some technical 
details that require a better understanding of general relativity, 
keeping only the details that are really necessary to understand 
the quantum aspects themselves.)
Different choices of time lead to different choices of the 
positive-frequency bases $\{ f_k(x) \}$, and thus to different 
creation and destruction operators $a_k^{\dagger}$ and $a_k$, 
respectively. If
\begin{eqnarray}\label{2phi}
& \hat{\phi}(x)=\displaystyle\sum_k 
\hat{a}_kf_k(x)+\hat{a}_k^{\dagger}f_k^*(x) , & \nonumber \\
& \hat{\phi}(x)=\displaystyle\sum_l \hat{\bar{a}}_l\bar{f}_l(x)+
\hat{\bar{a}}_l^{\dagger}\bar{f}_l^*(x)  &
\end{eqnarray} 
are two such expansions in the bases $\{ f_k(x) \}$
and $\{ \bar{f}_l(x) \}$, respectively, 
it is easy to show that the corresponding 
creation and destruction operators are related by a linear 
transformation 
\begin{eqnarray}\label{bogol}
& \hat{\bar{a}}_l = \displaystyle\sum_k \alpha_{lk} \hat{a}_k +
\beta^*_{lk} \hat{a}_k^{\dagger} , & \nonumber \\
& \hat{\bar{a}}_l^{\dagger} = \displaystyle\sum_k 
\alpha^*_{lk} \hat{a}_k^{\dagger} + \beta_{lk} \hat{a}_k ,
\end{eqnarray}
where 
\begin{equation}\label{bogcoef}
\alpha_{lk}\equiv (\bar{f}_l,f_k) , \;\;\;
\beta^*_{lk}\equiv (\bar{f}_l,f_k^*) ,
\end{equation}
are given by the scalar products defined as in (\ref{scalprod2}).
(To derive (\ref{bogol}), take the scalar product of both expressions
in (\ref{2phi}) with $\bar{f}_{l'}$ on the left and use
the orthonormality relations 
$(\bar{f}_{l'},\bar{f}_{l})=\delta_{l'l}$, 
$(\bar{f}_{l'},\bar{f}^*_{l})=0$.)
The transformation (\ref{bogol}) between the two sets of 
creation and destruction operators is called {\em Bogoliubov
transformation}. Since both bases are orthonormal, the Bogoliubov
coefficients (\ref{bogcoef}) satisfy
\begin{equation}\label{bogol2}
\sum_k (\alpha_{lk}\alpha^*_{l'k}-\beta^*_{lk}\beta_{l'k})=\delta_{ll'} ,
\end{equation}
where the negative sign is a consequence of the fact that 
negative frequency solutions have negative norms, i.e., 
$(f_k^*,f_{k'}^*)=-\delta_{kk'}$, 
$(\bar{f}_l^*,\bar{f}_{l'}^*)=-\delta_{ll'}$.
One can show that (\ref{bogol2}) provides that $\hat{\bar{a}}_l$ 
and $\hat{\bar{a}}_l^{\dagger}$ also satisfy the same commutation relations
(\ref{combos}) as $\hat{a}_k$ and $\hat{a}_k^{\dagger}$ do. 
A physically nontrivial Bogoliubov transformation 
is that in which at least some of the $\beta_{lk}$ coefficients 
are not zero. Two different definitions of the particle-number operators 
are
\begin{equation}
\hat{N}=\sum_k \hat{N}_k , \;\;\;
\hat{\bar{N}}=\sum_l \hat{\bar{N}}_l ,
\end{equation}
where
\begin{equation}
\hat{N}_k=\hat{a}_k^{\dagger}\hat{a}_k , \;\;\;
\hat{\bar{N}}_l=\hat{\bar{a}}_l^{\dagger}\hat{\bar{a}}_l .
\end{equation}
In particular, from (\ref{bogol}), it is easy to show that
the vacuum $|0\rangle$ having the property  
$\hat{N}|0\rangle=0$ has the property
\begin{equation}\label{parcreat}
\langle 0|\hat{\bar{N}}_l|0\rangle=\sum_k |\beta_{lk}|^2 .
\end{equation}
For a nontrivial Bogoliubov transformation,
this means that the average number of particles in the
no-particle state $|0\rangle$ is a 
state full of particles when the particles are defined by
$\hat{\bar{N}}$ instead of $\hat{N}$! Conversely, 
the no-particle state $|\bar{0}\rangle$ having the property 
$\hat{\bar{N}}|\bar{0}\rangle=0$ is a state full of particles
when the particles are defined with $\hat{N}$.
So, what is the right operator of the number of particles,
$\hat{N}$ or $\hat{\bar{N}}$? How to find the right 
operator of the number of particles? The fact is that, in general, 
a clear universally accepted answer to this question is not known! 
Instead, there are several possibilities that we discuss below.

One possibility is that the dependence of the particle concept 
on the choice of time means that the concept of particles 
depends on the observer. The best known example of this interpretation 
is the Unruh effect \cite{unruh,unruh2,birdav}, according to which 
a uniformly accelerating observer perceives the standard Minkowski vacuum
(defined with respect to time of an inertial observer in Minkowski flat 
spacetime) as a state with a huge number of particles with a thermal 
distribution of particle energies, with the temperature proportional 
to the acceleration. Indeed, this effect (not yet experimentally confirmed!)
can be obtained by two independent 
approaches. The first approach is by a Bogoliubov transformation as indicated
above, leading to \cite{unruh,birdav}
\begin{equation}\label{unruh}
\langle 0|\hat{\bar{N}}_l|0\rangle=\frac{1}{e^{2\pi\omega_l/a}-1},
\end{equation}
where $a$ is the proper acceleration perceived by the 
accelerating observer, $\omega_l$ is the frequency associated with 
the solution $\bar{f}_l(x)$, 
and we use units in which $\hbar=c=1$.
(Coordinates of a uniformly accelerating observer are known
as Rindler coordinates \cite{rind}, so the
quantization based on particles defined with respect 
to the Rindler time is called 
Rindler quantization.) We see that the right-hand side of (\ref{unruh}) looks 
just as a Bose-Einstein distribution at the temperature $T=a/2\pi$
(in units in which the Boltzmann constant is also taken to be unit).
The second approach is by studying the response of 
a theoretical model of an accelerating particle detector, 
using only the standard 
Minkowski quantization without the Bogoliubov transformation. 
However, these two approaches are {\em not}
equivalent \cite{padmun,nikolun}. Besides, such a dependence of particles 
on the observer is not relativistically covariant. In particular, it is not 
clear which of the definitions of particles, if any, acts as a source 
for a (covariantly transforming) gravitational field. 

An alternative 
is to describe particles in a unique covariant way in terms of local 
particle currents \cite{nikolcur}, but such an approach requires a 
unique choice of a preferred time coordinate. For example, for a hermitian 
scalar field (\ref{field}), the particle current is
\begin{equation}\label{enik1}
\hat{j}^{\rm P}_{\mu}(x)=i\hat{\psi}^{\dagger}(x) 
\!\stackrel{\leftrightarrow\;}{\partial_{\mu}}\! \hat{\psi}(x) ,
\end{equation}
which (unlike (\ref{cur}) with (\ref{e48})) requires the identification of 
the positive- and negative-frequency parts 
$\hat{\psi}(x)$ and $\hat{\psi}^{\dagger}(x)$, respectively. 
Noting that the quantization of fields 
themselves based on the functional Schr\"odinger equation 
(\ref{funcsch}) also requires a choice of a preferred time coordinate, 
it is possible that a preferred time coordinate emerges dynamically 
from some nonstandard covariant method of quantization, such as that in
\cite{nikddw}.

Another possibility is that the concept of particles as fundamental
objects simply does not make sense in QFT \cite{birdav,dav}.
Instead, all observables should be expressed in terms of local 
fields that do not require the artificial identification of the 
positive- and negative-frequency parts. For example, such an observable
is the Hamiltonian density (\ref{hamden}), which represents
the $T^0_0$-component of the covariant energy-momentum tensor
$T^{\mu}_{\nu}(x)$. Whereas such an approach is very natural from 
the theoretical point of view according to which QFT is nothing
but a quantum theory of fields, the problem is to reconcile 
it with the fact that the objects observed in high-energy 
experiments are -- particles.  

\subsection{Particle creation by a classical field}

When the classical metric $g_{\mu\nu}(x)$ has a nontrivial dependence on 
$x$, the Klein-Gordon equation (\ref{KG}) for the field 
$\phi(x)$ generalizes to
\begin{equation}
\left(
\frac{1}{\sqrt{|g|}} \partial_{\mu}\sqrt{|g|} g^{\mu\nu}\partial_{\nu}
+m^2 \right) \phi=0 ,
\end{equation}
where $g$ is the determinant of the matrix $g_{\mu\nu}$.
In particular, if the metric is time dependent, then
a solution $f_k(x)$ having a positive frequency at some initial time
$t_{\rm in}$ may behave as a superposition of positive- and 
negative-frequency solutions at some final time $t_{\rm fin}$. 
At the final time, 
the solutions that behave as positive-frequency ones are some other 
solutions $\bar{f}_l(x)$. In this case, it seems natural 
to define particles with the operator $\hat{N}$ at the initial time 
and with $\hat{\bar{N}}$ at the final time. If the time-independent state
in the Heisenberg picture is given by the ``vacuum" $|0\rangle$, 
then $\langle 0|\hat{N}|0\rangle=0$ denotes that there are no 
particles at $t_{\rm in}$, while
(\ref{parcreat}) can be interpreted as a consequence of an evolution
of the particle-number operator, so that (\ref{parcreat})
refers only to $t_{\rm fin}$. This is the essence of the 
mechanism of particle creation by a classical gravitational 
field. The best known example is particle creation by a
collapse of a black hole, known also as Hawking radiation 
\cite{hawk}. (For more details, see also the classic textbook 
\cite{birdav}, a review \cite{brout}, and a pedagogic review \cite{jacob}.)
Similarly to the Unruh effect (\ref{unruh}), the Hawking particles 
have the distribution 
\begin{equation}\label{unruhbh}
\langle 0|\hat{\bar{N}}_l|0\rangle=\frac{1}{e^{8\pi GM\omega_l}-1},
\end{equation}
where $G$ is the Newton gravitational constant which has a dimension
(energy)$^{-2}$ and $M$ is the black-hole mass.
This is the result obtained by defining particles with respect to
a specific time, that is, the 
time of an observer static with respect to the black hole and staying 
far from the black-hole horizon. 
Although (\ref{unruhbh}) looks exactly like a quantum 
Bose-Einstein thermal distribution at the temperature
\begin{equation}\label{Thawk}
T=\frac{1}{8\pi GM} ,
\end{equation}
this distribution is independent of the validity of the 
bosonic quantum commutation relations (\ref{combos}).
Instead, it turns out that 
the crucial ingredient leading to a thermal distribution
is the existence of the horizon \cite{padmrep},
which is a {\em classical} observer-dependent general-relativistic 
object existing not only for black holes, but also for 
accelerating observers in flat spacetime. 
Thus, the origin of this thermal distribution 
can be understood even with classical 
physics \cite{padmclas}, while only the mechanism of particle creation 
itself is intrinsically quantum. In the literature, 
the existence of thermal Hawking 
radiation often seems to be widely accepted as a fact. Nevertheless, since 
it is not yet experimentally confirmed and since it rests 
on the theoretically ambiguous concept of particles in curved spacetime
(the dependence on the choice of time), 
certain doubts on its existence are still reasonable (see, e.g.,
\cite{padmprl,nikolcur,bel} and references therein).
Thus, the existence of Hawking radiation can also be qualified as a myth.

The classical gravitational field is not the only classical 
field that seems to be able to cause a production of 
particles from the vacuum. 
The classical electric field seems to be able to produce 
particle-antiparticle pairs \cite{schw,manog,brout}, 
by a mechanism similar to the gravitational one. For discussions 
of theoretical ambiguities lying behind this theoretically predicted 
effect, see \cite{padmprl,nikolcur,nikolcrit}. 
  
\subsection{Particles, fields, or something else?}

Having in mind all these foundational problems with the concept of particle
in QFT, it is still impossible to clearly and definitely 
answer the question whether the 
world is made of particles or fields. Nevertheless, practically oriented 
physicists may not find this question disturbing as long as the 
formalism, no matter how incoherent it may appear to be, 
gives correct predictions on measurable quantities. 
Such a practical attitude seems to be justified by the 
vagueness of the concept of reality inherent to QM itself. 
Indeed, one can
adopt a hard version of orthodox interpretation of QM according 
to which information about reality is more fundamental
than reality itself and use it to justify  
the noncovariant dependence of particles (as well as some other 
quantities) on the observer \cite{peres}. However, in the standard 
orthodox QM, where rigorous no-hidden-variable theorems exist (see
Sec.~\ref{NOREAL}),
at least the operators are defined unambiguously.
Thus, even the hard-orthodox interpretation of 
QM is not sufficient to justify the interpretation of the  
particle-number-operator ambiguities as different realities 
perceived by different observers. An alternative to this 
orthodox approach is an objective-realism approach in which 
both particles and fields separately exist, which is a 
picture that seems to be particularly coherent in the 
Bohmian interpretation \cite{nikolpf}.    

Finally, there is a possibility that the world is made
neither of particles nor of fields, but of strings.
(For an excellent pedagogic introduction to string theory see 
\cite{zwie}. In particular, this book also breaks 
one myth in physics -- the myth that string theory is mathematically 
an extremely 
complicated theory that most other physicists
cannot easily understand. For a more concise pedagogic introduction
to string theory see also \cite{szabo}.) In fact, many string theorists
speak about the existence of strings as a definite fact. Fortunately,
there is still a sufficiently large number of authoritative physicists 
that are highly skeptical about string theory, which does not allow
string theory to become a widely accepted myth. 
Nevertheless, string theory possesses some remarkable 
theoretical properties that makes it
a promising candidate for a more fundamental description of nature.
According to this theory, particles are not really pointlike, 
but extended one-dimensional objects. Their characteristic length, 
however, is very short, 
which is why they appear as pointlike with respect to our 
current experimental abilities to probe short distances.
However, just as for particles, there is first quantization of 
strings, as well as second quantization that leads to string field theory.
Thus, even if string theory is correct, there is still 
a question whether the fundamental objects are strings or 
string fields. However, while first quantization of strings 
is well understood, string field theory is not. Moreover, 
there are indications that string field theory may {\em not} be
the correct approach to treat strings \cite{polc}.
Consequently, particles (that represent an approximation of strings) 
may be more fundamental than fields. Concerning the issue of 
objective reality, there are indications 
that the Bohmian interpretation of strings may be even more 
natural than that of particles \cite{nikolstr1,nikolstr3}. The Bohmian 
interpretation of strings also breaks some other myths 
inherent to string theory \cite{nikolstr2}.     

\section{Black-hole entropy is proportional to its surface}
\label{BH}

As this claim is not yet a part of standard textbooks,
this is not yet a true myth. Nevertheless, 
in the last 10 or 20 years
this claim has been so often repeated by experts in the field
(the claim itself is about 30 years old) 
that it is very likely that it will soon become a true myth.
Before it happens, let me warn the wider physics community
that this claim is actually very dubious. 

\subsection{Black-hole ``entropy" in classical gravity}

The claim in the title of this section
is actually a part of a more general belief that there exists 
a deep relation between black holes and thermodynamics. The first evidence 
supporting this belief came from certain classical properties of black holes
that, on the mathematical level, resemble the laws of thermodynamics 
\cite{beken,hawk2}. 
(For general pedagogic overviews, see, e.g., \cite{carrol,hawk3}
and for an advanced pedagogic review with many technical details, 
see \cite{town}.) Black holes are dynamical objects that can start
their evolution from a huge number of different initial states, 
but eventually end up in a highly-symmetric equilibrium stationary state 
specified only by a few global conserved physical quantities, 
such as their mass $M$ (i.e., energy $E$), electric charge $Q$, 
and angular momentum $J$. The physical laws governing the behavior
of such equilibrium black holes formally resemble the laws 
governing the behavior of systems in 
thermodynamic equilibrium. The well-known four laws of thermodynamics 
have the following black-hole analogues:
\begin{itemize}
\item
Zeroth law: There exists a local quantity called surface gravity 
$\kappa$ (which can be viewed as the general-relativistic analog 
of the Newton gravitational field $GM/r^2$)
that, in equilibrium, turns out to be constant everywhere    
on the black-hole horizon. This is an analog of temperature $T$ 
which is constant in thermodynamic equilibrium.
\item
First law: This is essentially the law of energy conservation, 
which, both in the black-hole and the thermodynamic case, has an 
origin in even more fundamental laws. As such, this analogy 
should not be surprising, but it is interesting that in both cases
the conservation of energy takes a mathematically similar form. 
For black holes it reads
\begin{equation}\label{dM}
dM=\frac{\kappa}{8\pi G}dA +\Omega dJ +\Phi dQ,
\end{equation}
where $A$ is the surface 
of the horizon, $\Omega$ is the angular velocity 
of the horizon, and $\Phi$ is the electrostatic potential
at the horizon. This is analogous to the thermodynamic first law
\begin{equation}\label{dE}
dE=TdS-pdV+\mu dN ,
\end{equation}
where $S$ is the entropy, $p$ is the pressure, $V$ is the volume,
$\mu$ is the chemical potential, and $N$ is the number of particles.
In particular, note that the black-hole analog of the entropy $S$
is a quantity proportional to the black-hole surface $A$.
This allows us to introduce the black-hole ``entropy"
\begin{equation}\label{S=A}
S_{\rm bh}=\alpha A ,
\end{equation}
where $\alpha$ is an unspecified constant.  
\item
Second law: Although the fundamental microscopic physical 
laws are time reversible, the macroscopic laws are not.
Instead, disorder tends to increase with time. In the 
thermodynamic case, it means that entropy cannot decrease with 
time, i.e., $dS\geq 0$. In the gravitational case, owing 
to the attractive nature of the gravitational force, it turns out 
that the black-hole surface cannot decrease with time, i.e., 
$dA\geq 0$.
\item
Third law: It turns out that, by a realistic 
physical process, it is impossible to reach the state with 
$\kappa=0$. This is analogous to the third law of thermodynamics 
according to which, by a realistic physical process, 
it is impossible to reach the state with $T=0$. 
\end{itemize}

Although the analogy as presented above is suggestive, it is clear 
that classical black-hole parameters are conceptually very different 
from the corresponding thermodynamic parameters.    
Indeed, the formal analogies above 
were not taken very seriously at the beginning. In particular, 
an ingredient that is missing for a full analogy between classical 
black holes and thermodynamic systems is -- radiation 
with a thermal spectrum. Classical black holes
(i.e., black holes described by the classical Einstein
equation of gravity)
do not produce radiation with a thermal spectrum.

\subsection{Black-hole ``entropy" in semiclassical gravity}

A true surprise happened when Hawking found out \cite{hawk} that
{\em semiclassical} (i.e., gravity is treated classically 
while matter is quantized) black holes  
not only radiate (which, by itself, 
is not a big surprise), but radiate exactly with a thermal spectrum 
at a temperature proportional to $\kappa$. In the special case of a 
black hole with $J=Q=0$, this temperature is equal to
(\ref{Thawk}). 
Since $dJ=dQ=0$, we attempt to write (\ref{dM}) as
 \begin{equation}\label{td1}
dS_{\rm bh}=\frac{dM}{T} ,
\end{equation}
which corresponds to (\ref{dE}) with $dV=dN=0$.
From (\ref{Thawk}), we see that
\begin{equation}\label{cons1}
\frac{dM}{T}=8\pi GMdM .
\end{equation}
From the Schwarzschild form of the black-hole metric in the 
polar spacial coordinates ($r,\vartheta,\varphi$) (see, e.g.,
\cite{carrol})
\begin{equation}
ds^2=\frac{dt^2}{1-\displaystyle\frac{2GM}{r}}
-\left( 1-\frac{2GM}{r} \right) dr^2 -r^2 (d\vartheta^2 +
{\rm sin}^2\vartheta \, d\varphi^2 ) ,
\end{equation}
we see that the horizon corresponding to the singular behavior 
of the metric is at the radius
\begin{equation}
r=2GM .
\end{equation}
Consequently, the surface of the horizon is equal to
\begin{equation}\label{bhpom}
A=4\pi r^2=16\pi G^2 M^2. 
\end{equation}
Therefore, (\ref{S=A}) implies
\begin{equation}\label{cons2}
dS_{\rm bh}=\alpha 32\pi G^2 MdM.
\end{equation}
Thus, we see that (\ref{cons1}) and (\ref{cons2}) are really 
consistent with (\ref{td1}), provided that $\alpha=1/4G$.
Therefore, (\ref{S=A}) becomes   
\begin{equation}\label{S=A2}
S_{\rm bh}=\frac{A}{4G} .
\end{equation}
In fact, (\ref{S=A2}) turns out to be a generally valid relation, 
for arbitrary $J$ and $Q$. 

Now, with the results (\ref{Thawk}) and (\ref{S=A2}), the analogy
between black holes and thermodynamics seems to be more complete.
Nevertheless, it is still only an analogy. Moreover, 
thermal radiation (which is a kinematical effect depending 
only on the metric) is not directly logically related to the four laws 
of classical black-hole ``thermodynamics" (for which the validity
of the dynamical Einstein equations is crucial) \cite{viss}. 
Still, many physicists
believe that such a striking analogy cannot be a pure formal coincidence.
Instead, they believe that there is some even deeper meaning of this 
analogy. In particular, as classical horizons hide information from 
observers, while the orthodox interpretation of QM suggests a 
fundamental role of information available to observers, it is   
believed that this could be a key to a deeper understanding
of the relation between relativity and quantum theory \cite{peres}. 
As the correct theory of quantum gravity is not yet 
known (for reviews of various approaches to quantum gravity, 
see \cite{carl,alvar}), there is a belief that this deeper 
meaning will be revealed one day when we better
understand quantum gravity. Although this belief may turn out 
to be true, at the moment there is no real proof that this 
necessarily must be so.  

A part of this belief is that (\ref{S=A2}) is {\em not} merely
a quantity {\em analogous} to entropy, but that it really {\em is}
the entropy. However, in standard statistical physics (from which 
thermodynamics can be derived), entropy is a quantity 
proportional to the number of the microscopic physical degrees of 
freedom. On the other hand, the derivation of (\ref{S=A2})
as sketched above does not provide a direct answer to the 
question what, if anything, these microscopic degrees of freedom are.
In particular, they cannot be simply the particles forming
the black hole, as there is no reason why the number of particles 
should be proportional to the surface $A$ of the black-hole boundary.
Indeed, as entropy is an extensive quantity, one expects 
that it should be proportional to the black-hole volume, rather 
than to its surface. It is believed that quantum gravity
will provide a more fundamental answer to the question
why the black-hole entropy is proportional to its surface, 
rather than to its volume.
Thus, the program of finding a microscopic derivation
of Eq.~(\ref{S=A2}) is sometimes referred to as  
``holly grail" of quantum gravity. (The expression ``holly grail"
fits nice with my expression ``myth".)    

\subsection{Other approaches to black-hole entropy}

Some results in quantum gravity already suggest a microscopic explanation
of the proportionality of the black-hole entropy with its surface.
For example, a loop representation of quantum-gravity kinematics
(for reviews, see, e.g., \cite{rov,rovbook})
leads to a finite value of the entropy of a surface, which coincides 
with (\ref{S=A2}) if one additional 
free parameter of the theory is adjusted appropriately. However, loop quantum
gravity does not provide a new answer to the question why the black-hole 
entropy should coincide with the entropy of its boundary. Instead, it uses
a classical argument for this, based on the observation that 
the degrees of freedom behind the horizon are invisible to outside 
observers, so that only the boundary of the black hole is relevant 
to physics observed by outside observers. (The book \cite{rovbook}
contains a nice pedagogic presentation of this classical argument.
Besides, it contains an excellent pedagogic presentation 
of the relational interpretation of general relativity, which, in particular,
may serve as a motivation for the conceptually much more dubious 
relational interpretation of QM \cite{rov1,rov2} mentioned in 
Sec.~\ref{NOREAL}.) Such an explanation of the black-hole entropy 
is not what is really searched for, as it does not completely 
support the four laws of black-hole ``thermodynamics", since 
the other extensive quantities such as mass $M$ and charge $Q$
contain information about the matter content of the {\em interior}.
What one wants to obtain is that the entropy of the {\em interior}
degrees of freedom is proportional to the boundary of the interior.

A theory that is closer to achieving this goal is string theory,
which, among other things, also contains a quantum theory of gravity.
Strings are one-dimensional objects containing an infinite 
number of degrees of freedom. However, not all degrees 
of freedom need to be excited. In low-energy states 
of strings, only a few degrees of freedom are excited, which corresponds 
to states that we perceive as standard particles. 
However, if the black-hole interior consists of one or a few 
self-gravitating 
strings in highly excited states, then the entropy associated 
with the microscopic string degrees of freedom is 
of the order of $GM^2$ (for reviews, see \cite{zwie,horow}).
This coincides with the semiclassical black-hole 
``entropy", as the latter is also of the order of 
$GM^2$, which can be seen from (\ref{S=A2}) and (\ref{bhpom}). 
The problem is that strings do not necessarily need to be 
in highly excited states, so the entropy of strings 
does not need to be of the order of $GM^2$. Indeed, the black-hole
interior may certainly contain a huge 
number of standard particles, which corresponds to a huge 
number of strings in low-excited states. It is not clear 
why the entropy should be proportional 
to the black-hole surface even then.

A possible reinterpretation of the relation (\ref{S=A2}) 
is that it does not necessarily denote the actual 
value of the black-hole entropy, but only the upper limit
of it. This idea evolved into a modern paradigm 
called {\em holographic principle} (see \cite{bousso} for a review), 
according to which the boundary of a region of space contains a lot 
of information about the region itself. However, a 
clear general physical explanation of the 
conjectured holographic principle, or that of the conjectured upper limit
on the entropy in a region, is still missing.   

Finally, let me mention that the famous 
black-hole entropy {\em paradox} that seems to suggest 
the destruction of entropy owing to the black-hole 
radiation (for pedagogic reviews, see \cite{gidd,stro}) 
is much easier to solve 
when $S_{\rm bh}$ is not interpreted as true entropy \cite{nikolbh}.  
Nevertheless, I will 
not further discuss it here as this paper is not about
quantum paradoxes (see, e.g., \cite{laloe}), but about quantum myths.

\section{Discussion and conclusion}

As we have seen, QM is full of ``myths", 
that is, claims that are often presented as definite facts, despite
the fact that the existing evidence supporting these claims
is not sufficient to proclaim them as true facts. 
To show that they are not true facts, I have also discussed the 
drawbacks of this evidence, as well as some alternatives.
In the paper, I have certainly not mentioned all myths existing
in QM, but I hope that I have catched the most famous
and most fundamental ones, appearing in several fundamental branches
of physics ranging from
nonrelativistic quantum mechanics of single particles to
quantum gravity and string theory.

The question that I attempt to answer now is -- why the myths in QM
are so numerous? Of course, one of the reasons is certainly the fact 
that we still do not completely understand QM at the most fundamental level.
However, this fact by itself does not explain why 
quantum physicists (who are supposed
to be exact scientists) are so tolerant and sloppy about arguments 
that are not really the proofs, thus allowing the myths to form. 
To find a deeper reason, let me first note that
the results collected and reviewed in this paper 
show that the source of disagreement among physicists on the validity 
of various myths is not of mathematical origin, but of 
conceptual one. However, in 
classical mechanics, which is well-understood
not only on the mathematical, but also on the conceptual level,
similar disagreement among physicists almost never occur.
Thus, the common origin of myths in QM must lie in the 
fundamental {\em conceptual difference between classical 
and quantum mechanics}. But, in my opinion,
the main conceptual difference between classical and quantum mechanics
that makes the latter less understood on the conceptual level 
is the fact that the former introduces a clear notion of objective reality 
even without measurements. (This is why I referred 
to the myth of Sec.~\ref{NOREAL} as the {\em central} myth in QM.) 
Thus, I conclude that {\em the main reason for the existence of myths in QM
is the fact that QM does not give a clear answer to the question 
what, if anything, objective reality is}.  

To support the conclusion above, let me illustrate it by a simple 
model of objective reality. Such a model may seem to be 
naive and unrealistic, or may be open to further refinements, 
but here its only purpose is to demonstrate how a
model with explicit objective reality immediately gives clear
unambiguous answers to the questions whether the myths discussed 
in this paper are true or not. The simple model of objective reality  
I discuss is a Bohmian-particle interpretation, according to which 
particles are objectively existing pointlike objects having deterministic 
trajectories guided by (also objectively existing) wave functions.
To make the notion of particles and their ``instantaneous" interactions
at a distance unique, I assume that there is a single preferred system of 
relativistic coordinates, roughly coinciding with the global system 
of coordinates with respect to which the cosmic microwave backround 
is homogeneous and isotropic. Now let me briefly consider the basic claims
of the titles of all sections of the paper.
Is there a wave-particle duality? Yes, because both particles and wave
functions 
objectively exist. Is there a time-energy uncertainty relation? No,
at least not at the fundamental level, because the theory is 
deterministic. Is nature fundamentally 
random? No, in the sense that both waves and particle trajectories satisfy 
deterministic equations. Is there reality besides the measured reality?
Yes, by the central assumption of the model. 
Is QM local or nonlocal? It is nonlocal,
as it is a hidden-variable theory consistent with standard 
statistical predictions of QM.
Is there a well-defined relativistic QM? Yes, because, by assumption, 
relativistic particle trajectories are well defined with the aid of a 
preferred system of coordinates. Does quantum field theory (QFT) 
solve the problems of relativistic QM? No, because particles are 
not less fundamental than fields. Is QFT a theory of particles?
Yes, because, by assumption, particles {\em are} fundamental
objects. (If the current version of QFT is not completely compatible 
with the fundamental notion of particles, then it is QFT that 
needs to be modified.) 
Is black-hole entropy proportional to its surface? To obtain 
a definite answer to this last question, I have to further specify 
my model of objective reality. For simplicity, I assume that
gravity is {\em not} quantized (currently known facts
do not actually exclude this possibility), but determined
by a classical-like equation that, at least at sufficiently large
distances, has the form of a classical Einstein equation in which ``matter" 
is determined by the actual particle positions and velocities. 
(For more details of such a model, see \cite{niksemicl}.)
In such a model,
the four laws of black-hole ``thermodynamics" are a direct consequence 
of the Einstein equation, and there is nothing to be explained 
about that. The quantity $S_{\rm bh}$ is only {\em analogous}
to entropy, so the answer to the last question is -- no.
Whatever (if anything) the true quantum mechanism of objective 
particle creation near the black-hole horizon might be 
(owing to the existence of a preferred time, the mechanism based 
on the Bogoliubov transformation seems viable), 
the classical properties of gravity near the horizon imply
that the distribution of particle energies will be thermal
far from the horizon, which also does not require an additional 
explanation and is {\em not} directly related to the 
four laws of black-hole thermodynamics \cite{viss}.

Of course, with a different model of objective reality, the answers
to some of the questions above may be different. But the point is 
that the answers are immediate and obvious. With 
a clear notion of objective reality, there is not much room for 
myths and speculations. It does not prove that objective reality 
exists, but suggests that this is a possibility that
should be considered more seriously. 

To conclude, the claim that the fundamental principles
of quantum theory are today completely understood, so that it only
remains to apply these principles to various practical physical problems --
is also a myth. Instead, quantum theory is a theory which is not yet
completely understood at the most fundamental level and is open
to further fundamental research. Through this paper, I have
demonstrated this by discussing various fundamental myths in QM for which
a true proof does not yet really exist. I have also demonstrated 
that all these myths are, in one way or another, related to the 
central myth in QM according to which objective unmeasured reality
does not exist. I hope that this review will contribute to a 
better general conceptual understanding of quantum theory and make readers 
more cautios and critical before accepting various claims 
on QM as definite facts.

\section*{Acknowledgments}
\addcontentsline{toc}{section}{Acknowledgments}

As this work comprises the foundational background for a large part 
of my own scientific research in several seemingly different branches 
of theoretical physics, it is impossible to name all my colleagues 
specialized in different branches of physics that indirectly 
influenced this work through numerous discussions and objections
that, in particular, helped me become more open minded by 
understanding how the known physical facts can be viewed 
and interpreted in many different inequivalent ways, without 
contradicting the facts themselves.
Therefore, I name only J. M. Karim\"aki who suggested 
concrete improvements of this paper itself.   
I am also grateful to the anonymous referees whose constructive critical
objections stimulated further improvements and clarifications
in the paper.
This work was supported by the Ministry of Science and Technology of the
Republic of Croatia. 
%under Contract No.~0098002.

\end{document}